\documentclass[a4paper,11pt]{article}
\pdfoutput=1 

\usepackage{jcappub} 

\usepackage[T1]{fontenc} 
\hypersetup{linkcolor=red,citecolor=green,filecolor=cyan,urlcolor=magenta}

\def\be{\begin{equation}}
\def\ee{\end{equation}}
\def\ba{\begin{eqnarray}}
\def\ea{\end{eqnarray}}
\usepackage{mathptmx}
\usepackage{indentfirst}
\usepackage{ulem}
\usepackage{amsmath,amssymb,latexsym,bm}
\usepackage{graphicx}    
\usepackage{amsmath}
\usepackage{kotex}
\usepackage{xcolor}
\renewcommand{\d}{\operatorname{d}}

\newcommand{\nn}{\nonumber}

\title{The Kaiser-Rocket effect:  three decades  and counting}

\author[a,b]{Benedict Bahr-Kalus}
\author[c,d,e]{Daniele Bertacca}
\author[b,f]{Licia Verde}
\author[g]{Alan Heavens}

\affiliation[a]{
Korea Astronomy and Space Science Institute, 776 Daedeok-daero, Yuseong-gu, Daejeon 34055, Republic of Korea.}
\affiliation[b]{
ICC, University of Barcelona, IEEC-UB, Mart\'i i Franqu\`es, 1, E-08028 Barcelona, Spain.}
\affiliation[c]{Dipartimento di Fisica e Astronomia ``G. Galilei'', Universit\'a degli Studi di Padova, via Marzolo 8, I-35131, Padova, Italy.}
\affiliation[d]{INFN Sezione di Padova,  I-35131 Padova, Italy.}
\affiliation[e]{INAF - Osservatorio Astronomico di Padova, Vicolo dell'Osservatorio 5, I-35122 Padova, Italy.}
\affiliation[f]{Instituci\'o Catalana de Recerca i Estudis Avan\c{c}ats, Passeig Lluis Companys 23, Barcelona 08010, Spain.}
\affiliation[g]{
Imperial Centre for Inference and Cosmology (ICIC), Imperial College, London SW7 2AZ, U.K.}

\emailAdd{benedictbahrkalus@kasi.re.kr}
\emailAdd{daniele.bertacca@pd.infn.it}
\emailAdd{liciaverde@icc.ub.edu}
\emailAdd{a.heavens@imperial.ac.uk}
\abstract{The peculiar motion of the observer, if not accurately  accounted for, is bound to induce a well-defined clustering signal in the distribution of galaxies. This signal is related to the Kaiser rocket effect. Here we examine the amplitude and form of this effect, both analytically and numerically, and discuss possible implications for the analysis and interpretation of forthcoming  cosmological surveys.
For an idealistic cosmic variance dominated full-sky survey with a Gaussian selection function peaked at $z\sim 1.5$ it is a $> 5\sigma$ effect and it can in principle bias very significantly the inference of cosmological parameters, especially for primordial non-Gaussianity.  For forthcoming surveys, with realistic masks and selection functions, the Kaiser rocket is not a significant concern for cosmological parameter inference except perhaps for primordial non-Gaussianity studies. However, it is a systematic effect,  whose origin, nature and imprint on galaxy maps are well known and thus should be subtracted or mitigated.  We present several approaches to do so.}

\begin{document}
\maketitle
\flushbottom

\section{Introduction}

Following the standard cosmological paradigm, 
and using the temperature maps of the Cosmic Microwave Background (CMB) radiation, 
it is possible to measure directly the dipole anisotropy and thus identify the bulk flow related to the motion of the Local Group (LG)  \cite{Juszkiewicz1990}. For example, Ref.~\cite{Bennett:2003bz} measured the LG velocity inferred from the CMB dipole (for completeness, see also \cite{Kogut:1993ag,Fixsen:1996nj,Hinshaw:2008kr}).
This is firmly grounded on the cosmological principle, assumes that the primordial dipole is negligible,  and implies that there is a preferred reference frame: the frame where the CMB looks statistically isotropic. Relaxing the assumption that the primordial dipole is negligible, our local motion correlates higher CMB multipoles due to modulation and aberration effects. A recent analysis of this effect has shown that our local motion is consistent with the CMB dipole, but is also consistent with up 33 per cent of the CMB dipole being caused by other effects \cite{Saha:2021bay}.

The local group's motion is the result of the cumulative gravitational pull of the surrounding distribution of matter in the Universe (e.g., \cite{Peebles1980}), motivating decades of efforts towards inferring the LG peculiar velocity from  measurements of the cosmological dipole in the galaxy distribution e.g.,
\cite{Yahil:1977zz, Yahil1980, Davis1982, Meiksin1986, Strauss1992, Schmoldt:1999cq, Kocevski:2005kr}.
The dipole due to our motion with respect to a rest frame where the galaxy distribution is statistically isotropic is expected to converge to the CMB  kinematic dipole if the galaxy survey is deep enough.
Ref.~\cite{Erdogdu:2005wi} finds that for the  2MRS survey the dipole velocity of the local group seems to converge to the CMB velocity dipole by $60$ Mpc$/h$. However, Ref.~\cite{Lavaux:2008th}  using the same catalogue finds that a depth of $100$ Mpc$/h$ is needed to agree with WMAP5 results for the CMB dipole at the 1 to 2-sigma level. 
A recent analysis, \cite{Nusser:2014sha}, concluded that the CMB frame can be gradually reached but the LG motion cannot be recovered to better than $150-200$ km s$^{-1}$ in amplitude and within an error of  $\simeq 10^{\circ}$ in direction (see also \cite{Yousuke2010}).

The effect of the peculiar motion of the observer on galaxy redshift surveys has recently attracted renewed attention, as it could affect the interpretation of the redshift itself~\cite{Davis:1907.12639, Glanville2011.04210}. While surprisingly small redshift errors can have a significant impact on cosmological inference, that induced by the LG motion turns out not to be a serious concern. However, the LG motion can produce a spurious effect on clustering measurements of the Universe's  Large Scale Structure (LSS) when traced by the galaxy distribution. It is related to the  {\it Rocket Effect}  \cite{Kaiser:1987qv},\footnote{This in turn is related to the effect described by Ellis and Baldwin \cite{Ellis:1984} for radio surveys.} where the local group motion can induce a spurious apparent overdensity in the direction of motion, which then may appear to be the cause of the motion in the first place.
In principle, the Kaiser Rocket effect should not be neglected \cite{Nusser:2014sha} and can be  corrected if the selection function is
sufficiently well-known \cite{Strauss:1995fz}. 
A recent investigation on the wide-angle correlations to the galaxy power spectrum in redshift space, \cite{Bertacca:2019wyg}, suggested that the Kaiser Rocket effect could dominate the local signal of the 2-point correlation function of galaxies at very large scales.
As it has already been pointed out in \cite{Hamilton:1995px}, the simplest way to analyse the {\it Rocket Effect} is to use the known value of  CMB dipole, i.e., to work directly in the CMB frame, e.g., see \cite{Yousuke2010}
(see also \cite{Taylor:1999ah} where they derived the observer's dipole directly from a redshift survey, without having to fully reconstruct the density field, both in the CMB and in the rest frames of the local group).
Finally, during the  last stage of our work,  ref. \cite{Castorina:2021xzs}  has appeared where the authors devote a long discussion and present  an analytical approach to the Rocket effect on the large-scale  monopole of the power spectrum due to the observer's velocity.

Motivated by this finding, here we quantify the Kaiser rocket effect and assess its role as a possible source of systematic effects in the estimation of cosmological parameters. Of particular concern is a possible systematic bias on the local non-Gaussianity parameter, $f_\mathrm{NL}$, inferred through the so-called halo bias effect. The $f_\mathrm{NL}$ signal appears as a large-scale excess power in the power spectrum/2-point correlation function of biased tracers, and it is these large scales that are most affected by the Kaiser-Rocket-induced spurious clustering signal. Indeed, the related Ellis \& Baldwin effect \cite{Ellis:1984} allows to interpret the large-scale clustering signal seen by the NRAO VLA Sky Survey (NVSS) as either a positive $f_\mathrm{NL}$ detection \cite{Xia:2010pe} or as an unexpectedly large kinetic dipole \cite{Chen:2015wga}.

The (Cartesian Fourier-based) power spectrum is one of the workhorse summary statistics in large-scale structure analyses. While working in Fourier space has several well-known advantages, including a direct and transparent correspondence with theory predictions, it has the disadvantage of losing spatial localisation. Because of this, the spurious signal of the  Kaiser Rocket effect is expected not to be obvious in the galaxy power spectrum.

The rest of the paper is organised as follows. 
In Section~\ref{sec:meth} we model the Kaiser Rocket signal as a contaminant for the power spectrum and illustrate its implementation on mock galaxy catalogues. If this signal is not accounted for, it can bias the inference of key cosmological parameters: we quantify this bias in Section~\ref{sec:fisherbias}.  We discuss recommendations for mitigating the effect from realistic surveys and present our conclusions in Section~\ref{sec:conclusions}.  

We present this work in natural units, i.e., we set the speed of light to unity.

\section{The Signal as a Power Spectrum Contaminant}
\label{sec:meth}
As shown in the original paper \cite{Kaiser:1987qv}, the Kaiser Rocket effect is a galaxy number density modulation with a dipolar pattern.  
Let us begin, 
as a first approximation, by assuming  that the rocket-effect-induced number-density modulation is a completely independent physical process from the cosmological clustering. A correlation is expected to arise from the fact that the density field itself sources the velocity of the observer.  However,  the contribution of the correlation to the monopole of the galaxy power spectrum is expected to be small. 
We shall confirm the validity of this assumption in Section \ref{sec:crosstest}. 
In the absence of correlations between the cosmological fluctuation field and the number density modulation due to the rocket effect, the power spectrum of the superposition of the two effects is just the sum of the two power spectra: $P(k)=P_\mathrm{cosmo}(k)+ P_\mathrm{rocket}(k)$. Hence, throughout most of this work, we consider the rocket power spectrum separately from the cosmological power spectrum. This separate treatment allows us to model $P_\mathrm{rocket}(k)$ starting from random catalogues, provided the number density is high enough so that the shot noise is not a problem. The effect of the window can also be studied on random catalogues. Therefore in what follows, unless otherwise stated, our "mocks" will be generated from random catalogues (i.e., no intrinsic clustering), by displacing the objects according to the motion of the observer and then applying the suitable window function. In the rest of this section,  first, we introduce an analytic model of the rocket power spectrum. Then we describe how we generate our random catalogues and how we displace the objects therein to obtain "mock" realisations that, in the following subsections, we compare to the model prediction, and we use to estimate $P_\mathrm{rocket}$ in the presence of a survey window. 

\subsection{Analytic Model of the Rocket Power Spectrum}
\label{sec:model}
Let us begin by recalling  that if the observer is in motion with velocity $v$ with respect to the CMB then the observed redshift $z$ is related to the cosmological redshift $\bar z$ by \cite{Davis:1907.12639}
\begin{equation}
    z = \bar z - (1 + \bar z) v_{\parallel 0} = \bar z - (1 + \bar z) v\cos\theta\,
    \label{eq:z_shift}
\end{equation}
where we call ${v}_{\parallel 0}$ the observer's velocity component aligned with their line-of-sight (LOS) and $\theta$ the angle between the LOS and the observer's motion vector.
Hence, the Kaiser rocket effect due to  a velocity ($v$, in natural units) is a LOS dependent shift to the mean number count $\bar N(z, \Omega)$.\footnote{What we call $\bar N(z, \Omega)$ is often called ${\d N(z)}/{d\Omega\d z}$ We chose our notation to avoid confusion with derivatives in this section.} at redshift $z$ and angular position $\Omega$. As the total number of objects is the same in both the observer's and  the CMB rest frame, we know that the observer's apparent number count $\hat N(z, \Omega)$ is related to the one in the CMB rest frame  by \begin{equation} \hat N(z, \Omega)\d z = \bar N(\bar z, \Omega)\d\bar z\;.
\end{equation}
At cosmological redshifts where $$\left\vert\bar z - z\right\vert\ll 1\, ,$$  at first order, we can write
\begin{equation}
    \hat N(z,\Omega) = \frac{\bar N(z,\Omega)+\frac{\d\bar N}{\d\bar z}\Big|_{\bar z=z}\left(\bar z - z\right) + \mathcal{O}\left[\left(\bar z - z\right)^2\right]}{1 - v\cos\theta}
    \label{eq:nbar_dipole}
\end{equation}
where $\theta$ is the angle between the LOS and the dipole direction. Writing this as a density contrast with respect to the unshifted $\bar N(z,\Omega)$, we obtain 
\begin{equation}
  \delta_\mathrm{rocket}(z,\theta)=\frac{\hat N( z,\Omega)-\bar N(z,\Omega)}{\bar N(z, \Omega)}=v\cos\theta + (1+z)\frac{\d \ln\bar N(z)}{\d z}v\cos\theta
  \label{eq:delta_rocket_z}
\end{equation}
at leading order.\footnote{From now on, we keep this expression only at leading order. The effect is small enough so this is a sufficiently good approximation.}
This is a dipolar modulation whose amplitude is driven by the redshift dependence of the mean number count and the velocity $v$. Note that the true underlying $\bar N(z,\Omega)$ is usually unknown and therefore typically estimated from the data as an average of the observed number density over the survey area:
\begin{equation}
    \hat{\bar N}(z) = \frac{1}{\Delta\Omega}\int_\mathrm{survey}\hat N( z,\Omega)\d\Omega = \bar N(z)\left(1 + v \langle \cos\theta\rangle\right) + \langle \cos\theta\rangle \frac{\d\bar N}{\d\bar z}\Big|_{\bar z=z}(1+z)v,
\end{equation}
where the angle brackets denote averaging over the survey area. The density contrast with respect to the estimated $\hat{\bar N}(z)$ is then 
\begin{equation}
    \hat\delta_\mathrm{rocket}(z,\theta)=v\left[\cos\theta - \langle\cos\theta\rangle\right]\left[ 1 + (1+z)\frac{\d \ln\bar N(z)}{\d\bar z}\right].
    \label{eq:delta_rocket_z_partial_sky}
\end{equation}
If we define ${\cal N}(z)\equiv \d\ln \bar{N}/\d\ln(1+z)$, 
considering that the cosmological redshift and comoving distance $s$ are related by the Hubble-Lema\^itre parameter, i.e. $\d s/ \d z=1/H(s(z)) \equiv 1/H(s)$, we can write eq. \eqref{eq:delta_rocket_z} entirely in terms of spherical coordinates:
\begin{equation}
    \hat\delta_\mathrm{rocket}(\mathbf{s})=v\cos\theta \left[ 1  + \frac{{\cal N}(s)}{H(s)}\right],
    \label{eq:delta_rocket_distance}
\end{equation}
where ${\cal N}(s)\equiv (1+z(s)) \d\ln \bar N/\d s$.
Note that $\theta$ denotes both the angle between the LOS and the dipole, as well as the azimuth of our coordinate system, by implicitly choosing the dipole as the azimuth reference point.\footnote{This gives rise to a signal which can be expressed by a single spherical harmonic coefficient (see below). 
}
This is valid without loss of generality as long as we consider the rocket effect on a full-sky survey.

It should be clear from equations \eqref{eq:delta_rocket_z} and \eqref{eq:delta_rocket_distance} that the natural basis to describe this effect is the so-called Spherical Fourier-Bessel basis (Fourier-Bessel decomposition/transform, see \cite{HT95, Lahav:1993, HT97}).
\begin{eqnarray}
\delta^{\mathrm{rocket}}_{\ell m}(k_z)&=&\sqrt{\frac{2}{\pi}}k_z v \int\cos\theta\left[1 + \frac{d\ln \bar{N}(z)}{d\ln(1+z)}\right] j_\ell(k_zz)Y^*_{\ell\,m}(\theta,\phi)z^2 \d z \d\Omega \\ \nn
&=&2\sqrt{\frac{2}{3}}k_z v \int\left[1 + \frac{d\ln \bar{N}(z)}{d\ln(1+z)}\right]\cos\theta j_1(k_zz)z^2dz \delta^K_{0m}\delta^K_{1\ell}
\label{eq:rocketFB}
\end{eqnarray}
    where  explicitly writing a $z$ subscript on $k$ we have specified that, if redshift is not transformed into distance the $k$ involved is dimensionless.\footnote{This formalism has been extended by \cite{Yoo:2013tc} and \cite{Bertacca:2017dzm} including also general relativistic effects.  Note that this decomposition is convenient for transformation of observed data, but the basis functions are not eigenfunctions of the Laplacian, which they are if $z$ is replaced by $r$ in flat space.}
Replacing the redshift $z$ with the comoving distance $s$, we can also  write 
\begin{equation}
\delta^{\rm Rocket}_{\ell m}(k)=2\sqrt{\frac{2}{3}}kv\int \cos\theta \left[ 1  + \frac{{\cal N}(s)}{H(s)}\right] j_1(ks)s^2 ds\delta^K_{0m}\delta^K_{1l}
\label{eq:rocketFB2}
\end{equation}

On the full sky, in Eqs.~\eqref{eq:rocketFB}, only the dipole component $\delta_{10}$ survives, if we identify the polar axis of the coordinate system with the direction of the residual dipole.
The presence of an angular mask induces non-zero higher multipoles:
\begin{equation}
\delta^{\rm rocket}_{\ell m}(k)=\sum _{\ell'\, m'}W_{\ell \ell'}^{m m'}\delta^{\rm Rocket}_{\ell' m'}(k)=W_{\ell 1}^{m 0}\delta^{\rm Rocket}_{10}(k)
\end{equation}
 where
 \begin{equation}
W_{\ell \ell'}^{m m'}=\int_{4\pi}Y_{\ell'}^{m'}(\Omega)M(\Omega)Y_\ell^{* m}(\Omega)d\Omega\,,
 \end{equation}
and $M(\Omega)$ denotes the angular mask.

Of course, one is not forced to select the direction of the dipole to  be the azimuth of our coordinate system, but this generic case is related to the equation above by a rotation matrix  (or Wigner $D$ matrices in harmonic space),
\begin{equation}
\delta^{{\rm Rocket}, \alpha,\beta,\gamma}_{\ell m}(k) =  \sum_{m'}  D^\ell_{m m'}(\alpha,\beta,\gamma)W_{\ell 1}^{m' 0} \delta^{\rm Rocket}_{10}\equiv \hat{W}_{\ell 1}^{m0}\delta^{\rm Rocket}_{10}
\end{equation}
where the Euler angles $\alpha, \beta,\gamma$ denote the rotation between the chosen coordinate system and one in which the direction of the dipole is the azimuth.

Despite several remarkable efforts (see e.g., \cite{Wang:2020wsx}), this is not the usual basis and approach for forthcoming galaxy surveys; the standard approach still prefers to use Cartesian coordinates and, consequently, a Fourier basis. The summary statistic of choice is the  (Cartesian) three-dimensional power spectrum. Accounting for a rocket effect dipolar signal using spherical Bessel basis would be much more transparent than using Fourier transforms and 3D power spectra, but it would significantly complicate the interpretation of the cosmological signal especially at scales significantly smaller than the survey size. 
 In the rest of this paper therefore we work with the  (Cartesian) 3D power spectrum.
 
To estimate the  contribution to the power spectrum from the rocket effect, we have to take the Fourier transform of $\delta_\mathrm{rocket}(\mathbf{s})$ over a volume $V$ with radius $R(V)$:
\begin{equation}
    \delta_\mathrm{rocket}(\mathbf{k})\equiv \sqrt{\frac{(2\pi)^{3}}{V}}\int_0^{2\pi}\int_{-1}^1\int_0^{R(V)} \delta_\mathrm{rocket}(\mathbf{s})e^{i\mathbf{k\cdot s}}s^2\d s\d\cos\theta\d\varphi.
\end{equation}
As $\delta_\mathrm{rocket}(\mathbf{s})$ does not depend on $\varphi$, we can evaluate the $\varphi$-integral using the integral representation of the modified Bessel function
\begin{equation}
    \int_0^{2\pi}e^{is\sin\theta(k_1\cos\varphi+k_2\sin\varphi)}\d\varphi=2\pi J_0(s\kappa\sin\theta),
\end{equation}
where $\kappa\equiv\sqrt{k_1^2+k_2^2}$ and $k_1, k_2, k_3$ are the wavevector components in cartesian coordinates. Since it is common-place in redshift space distortions literature that $k_3$ is aligned with the LOS, we want to emphasise here that in this work, $k_3$ points into the direction of the dipole. The remaining two integrals read
\begin{equation}
    \delta_\mathrm{rocket}(\kappa,k_3)=\sqrt{\frac{(2\pi)^{5}}{V}} v\int_{-1}^1\int_0^{R(V)} \cos\theta \left[ 1  + \frac{{\cal N}(s)}{H(s)}\right] J_0(s\kappa\sin\theta)e^{ik_3s\cos\theta}s^2\d s\d\cos\theta.
    \label{eq:deltarocketz}
\end{equation}

Defining $\mu \equiv k_3/k$, we can calculate the power spectrum monopole as
\begin{equation}
    P_\mathrm{rocket}(k) = \frac{1}{2}\int_{-1}^1\d\mu\left\langle\left\vert \delta_\mathrm{rocket}\left(k\sqrt{1-\mu^2}, k\mu\right) \right\vert^2\right\rangle.\\
    \label{eq:Pk_rocket}
\end{equation}

\subsection{Caveats and Limitations}

There are additional effects beyond the one on redshifts described in \S \ref{sec:model}. 
These are: the aberration effect, the possible discrepancy between the velocity ${\bf v}$ (related to  the local group velocity) and the  actual velocity of the observer, and the magnification bias.

In general, using a GR analysis, the square parenthesis in eq. \eqref{eq:delta_rocket_z_partial_sky} can be written as (e.g., see \cite{Maartens:2017qoa, Bertacca:2019wyg, Nadolny:2021hti})
\begin{equation}
\left[1+f_{\rm aber}+ (1+z)\Bigg[\frac{\partial \ln\bar N(z, L>L_{\rm min})}{\partial z}\Bigg]_{L_{\rm min}}-\frac{2\mathcal{Q}}{ {s}}\frac{(1+ {z})}{H}\right]\;,
\label{eq:GR}
\end{equation}
where
$L$ is the object's luminosity  and $\bar{N}(z,L>L_{\rm min})$ denotes $\bar{N}$ for all objects with luminosity $L$ above the minimal luminosity $L_{\rm min}$  at redshift z; , $\mathcal{Q}$ is the magnification bias\footnote{In general, the magnification bias is defined as $\mathcal{Q}=5 {\cal S}/2$, where 
\[{\cal S} =-{2\over 5}{\partial \ln \bar N(z,L>L_{\rm min})\over \partial \ln L_{\rm min}}\;.\]}
 and $f_{\rm aber}$ is the aberration effect.\footnote{Note that 
 \begin{equation}
 \label{ncom}
     \bar N(z, L>L_{\rm min})=s^2\, n_{\rm com}(z, L>L_{\rm min})/H(z)\;, 
 \end{equation}
 where $n_{\rm com}$ is the usual 3D comoving number density. Precisely, we define the comoving number density as $n_{\rm com}(z, L>L_{\rm min})=\int_{L_{\rm min}}^\infty  \phi(z, L)\,\d L$, where $\phi(z, L)$ is the usual bivariate luminosity function  which gives the comoving number density of galaxies per unit interval in redshift and luminosity (e.g see \cite{Broadhurst:1994qu}) In this case, in order to recover  results obtained in  \cite{Maartens:2017qoa, Bertacca:2019wyg, Nadolny:2021hti}, we have to set  $f_{\rm aber}=2$ and express $\bar N$ as in
 eq.~(\ref{ncom}).} As we now show, the first  term and part of the third term in the square bracket of eq.~(\ref{eq:GR}) appear already in eq.(\ref{eq:delta_rocket_z_partial_sky}), the second and forth 
 are additional effects discussed in this section. The aberration is due to the observer's velocity and changes the solid angle by 
\[ \left| \frac{\d \bar\Omega}{ \d \Omega} \right|= 1+2 {\bf v}_{\rm aber} \cdot {\bf n}\;, \]
where 
${\bf v}_{\rm aber}$ is the full velocity vector of the observer (with respect to a reference frame where the galaxy distribution is isotropic): this is composed by the velocity of the local group, the velocity of the galaxy in the local group, the velocity of the sun around the galaxy, the velocity of the Earth around the Sun and the velocity of the observer with respect to the Earth.

Following the discussion in \cite{Bertacca:2014hwa},  note that, for luminosity limited samples, 
\[
(1+z)\frac{\d \ln\bar N(z)}{\d z} = -2\mathcal{Q} \left[1+\frac{(1+ {z})}{s H}\right] +  (1+z)
\Bigg[\frac{\partial \ln\bar N(z, L>L_{\rm min})}
{\partial z}\Bigg]_{L_{\rm min}}\;.
\]
In summary, the square brackets in
{Eqs.~(\ref{eq:delta_rocket_z_partial_sky}) and  (\ref{eq:delta_rocket_distance})} should include the following additive term
\begin{equation}
    F:= f_{\rm aber}   + 2\mathcal{Q} \;.
    \label{eq:F}
\end{equation}
whose magnitude, however, is highly uncertain.
  If the velocity ${\bf v}_{\rm aber}$ is that of the local group then $f_{\rm aber}= 2$. However, in practice, the relevant velocity for the aberration term, $f_{\rm aber}$, is the total velocity vector, which includes also  a dominant (negative) correction due to the velocity of the Sun around the galaxy, ${\bf v}_\odot$, effectively reducing the amplitude of the aberration velocity vector by roughly 50\%. In reality, ${\bf v}_\odot$ is not perfectly  (anti-) parallel to ${\bf v}_{\rm loc}$ therefore $f_{\rm aber}$ could be a function of a residual angle between the two velocities. 
Neglecting this  extra modulation effect, we naively estimate that   $f_{\rm aber}$ get reduced by roughly 50\% bringing $f_{\rm aber} \simeq 1$.

The magnification bias arises for luminosity limited samples, and is also very dependent on the adopted galaxy sample and might have a marked redshift dependence too; its value cannot easily be predicted, but simulations indicate that it might be as large as $\sim 2$. 

Thus, based on the above discussion, let us highlight the following  consideration.
 When observers take data, these are  usually  corrected  --in redshift only-- for the change  induced  by the motion of the observatory and the Sun around the galaxy, but not angles or fluxes; these are usually  left untouched (but this might be observation- or survey- dependent). In general,  the dominant correction to the cosmological aberration is the Sun’s motion ${\bf v}_{\odot}$. The aberration effect is very important, about the same order of magnitude as the first and third term in eq.~(\ref{eq:F})  and cannot be ignored.  Observatories should therefore  correct not just the redshift  (as they do already) but also angles and fluxes  for the known motion (observatory, Earth and Sun). At this time however, one cannot fully model  the amplitude of this correction, in detail.
 
 In what follows we ignore the luminosity bias (as it is quite common to do), this is a different physical effect  which depends strongly on the type of objects selected; hence we leave  this for future work. Moreover we conservatively set $F=0$ by also suppressing $f_{\rm aber}$. It is important to note that the $N(z)$-independent factor in the dipole signal is highly uncertain but is expected to be at least of order unity.
 With this consideration  and these caveats in mind, we believe that the S/N calculations done in this paper are still a useful estimate.
 
\subsection{Implementation of Mocks}
\label{sec:mocks}

For most realistic survey configurations, eq. \eqref{eq:Pk_rocket} cannot be calculated analytically. For that reason, our approach is first to generate idealised mock catalogues to study the effect of the Kaiser rocket effect on galaxy surveys and to create more realistic catalogues later. The first step in doing so is generating  a random catalogue, with redshifts drawn from a given  $\bar N(z)$.

The Kaiser rocket effect adds a velocity component to the redshift-to-configuration-space relation (e.g. see \cite{Bertacca:2019wyg}).
From eq.~\eqref{eq:z_shift}, which defines $z$ and $\bar{z}$,
for a given cosmological model with Hubble-Lemaître  expansion rate $H(z)$, we can convert  the redshift correction into a distance shift 
\begin{equation}
    \bar s - {s} = \int_{{z}}^{\bar z}\frac{\d z'}{H(z')}.
\end{equation}
To simulate the Kaiser rocket effect on surveys and quantify its magnitude, we generate pairs of mock catalogues where one element of the pair is the true random catalogue (no intrinsic clustering) and the other one, the rocket mock, is obtained from the first using eq. \eqref{eq:z_shift} to replace the redshift $\bar z$ with the redshift $z$.

We obtain the power spectrum signal caused by the Kaiser rocket effect using the Feldman-Kaiser-Peacock estimator \cite[FKP]{Feldman:1993ky} where the signal is given by the rocket mock. Before using this machinery to forecast the effect for real cosmological surveys, we apply it to two toy models, comparing them to our analytic model (cf. Section \ref{sec:model}), and quantifying the expected magnitude of the effect and possible detectability.

\subsection{Toy Model I: A Full-Sky Survey with a Gaussian Radial Selection}
\label{sec:Gaussian_toy}

\begin{figure}
    \centering
    \includegraphics[width=\textwidth]{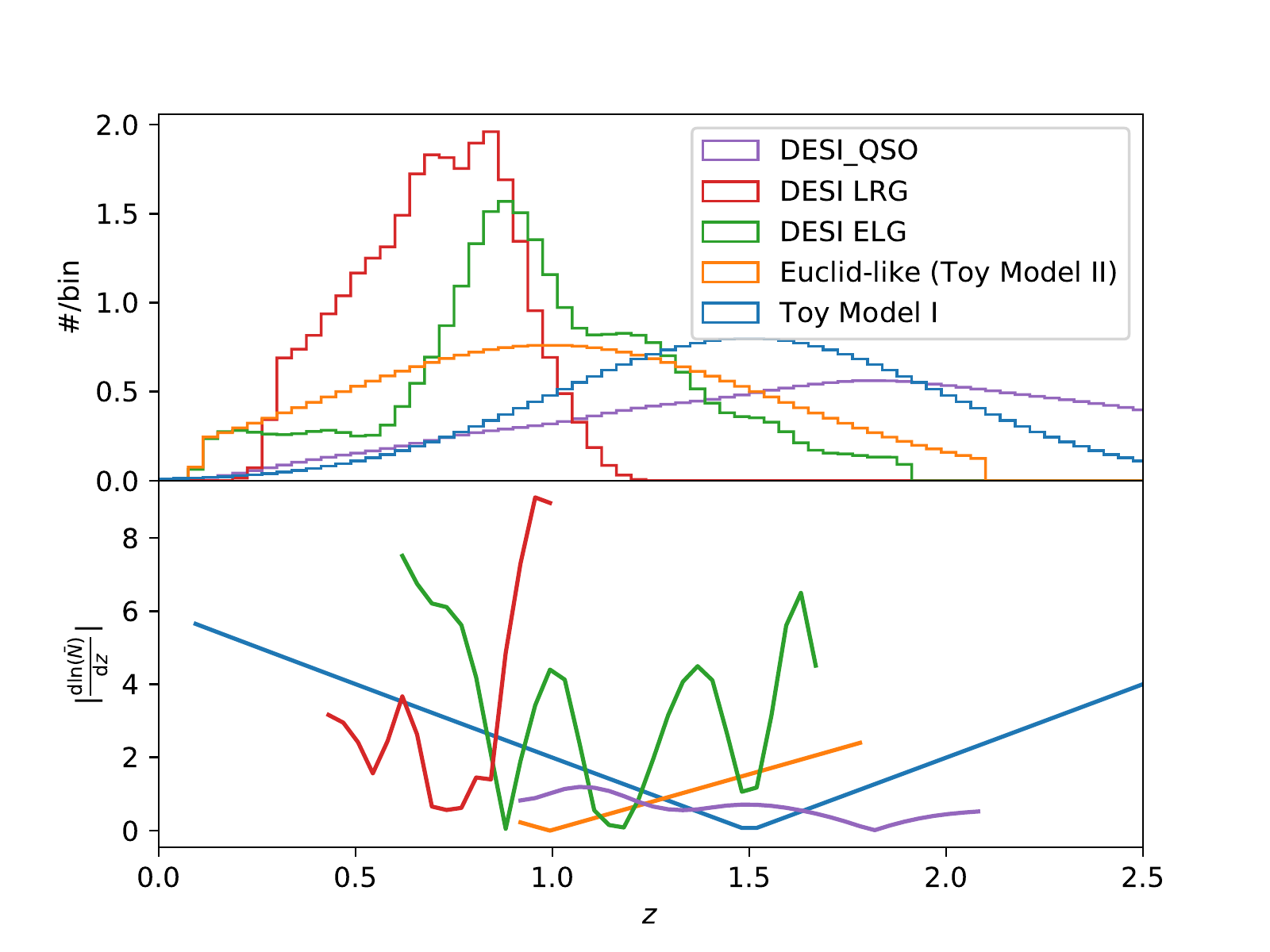}
    \caption{The number of objects per redshift bin in the random catalogues used throughout this study (top panel). For better comparison, and as the total number of objects is irrelevant for the Kaiser rocket effect, the histograms are normalised such that their integral is unity. In the bottom panel, we show the absolute value of the log-derivative of the selection function. We plot the log-derivatives only within the redshift range covered by the respective surveys. Note that the Euclid-like selection function is only a simple model whereas the DESI selection functions are based on its preliminary target selection. The DESI selection function might change during science validation though.}
    \label{fig:dNdz_histograms}
\end{figure}

For our first test, we consider a full-sky survey and an analytic expression for the radial selection function, which allows us to make some calculations of the rocket effect analytically.  We model the radial selection function as a Gaussian
\begin{equation}
    \bar N(z) \propto \exp\left(-\frac{1}{2}\left[\frac{z - 1.5}{0.5}\right]^2\right).
    \label{eq:Gaussian_n_of_z}
\end{equation}
The amplitude of $\bar N (z)$ is irrelevant because it is common to oversample the random catalogue to bring down the shot noise and because $\bar N (z)$ enters our analytic model in a log-derivative. We draw angular positions and redshifts in the range $0.1<z<3$ for $2.1 \times 10^8$ objects. This number yields, on average, 100 objects per grid cell in a cubic grid with 128 cells per dimension. We shift the random objects according to eq. \eqref{eq:z_shift}, where we set  the rocket velocity ${\bf v}$ to the CMB dipole amplitude measured by the Planck satellite \cite{Akrami:2018vks} $v=(1.23357 \pm 0.00036) \times 10^{-3}$. As directions do not matter as long as we observe the full sky, we choose a coordinate system aligned with the dipole direction.

\begin{figure}[!h]
    \centering
    \includegraphics[width=\textwidth]{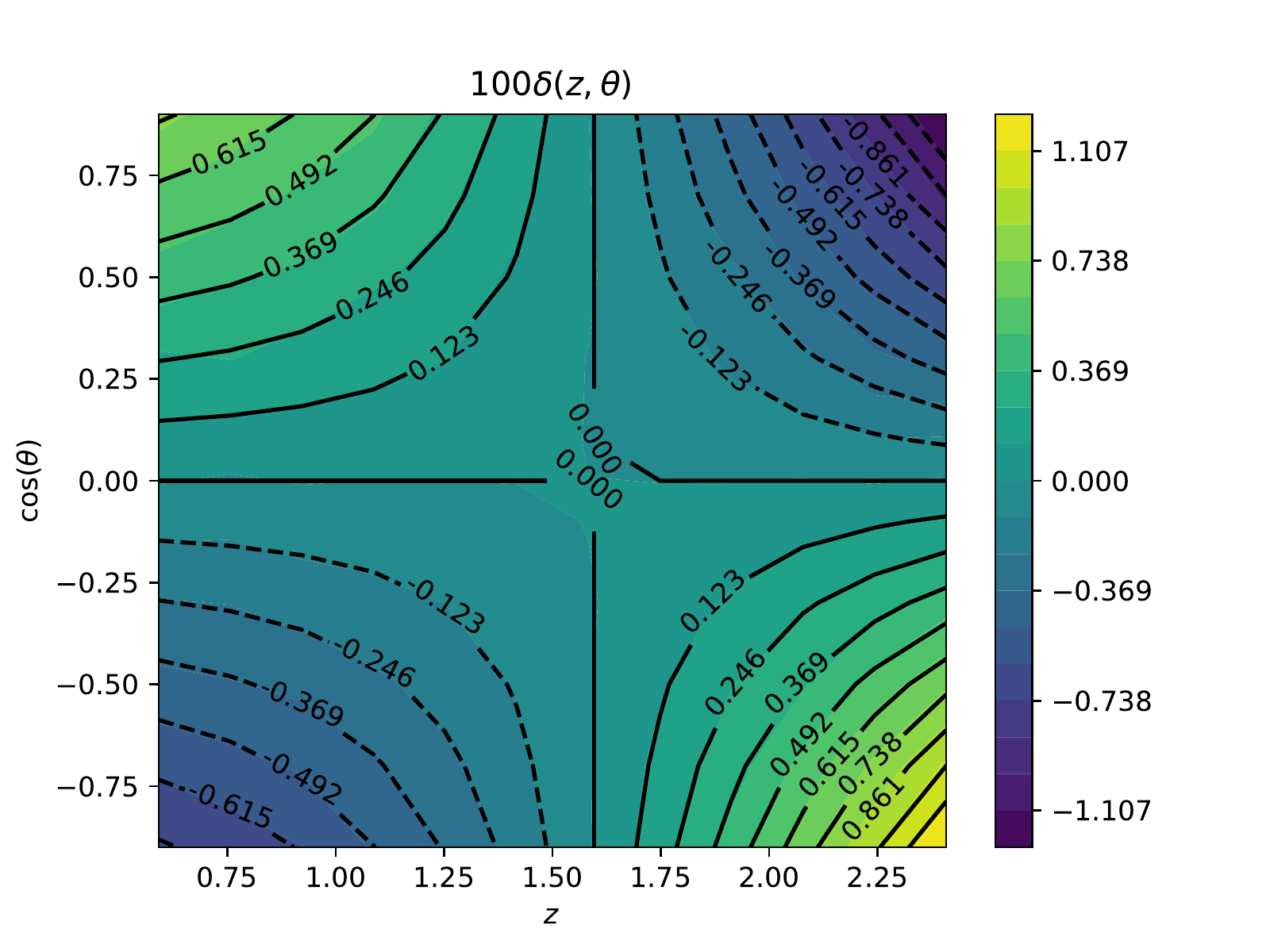}
    \caption{The density field $\delta_\mathrm{rocket}(z,\theta)$ as a function of redshift $z$ and angle $\theta$ between the dipole direction and the line of sight obtained for a mock realization (see text for more details). The black contours are predicted using eq. \eqref{eq:delta_rocket_z}. For clarity, we have multiplied all values of $\delta$ by 100. Note that we bin the mock catalogue in $\cos(\theta)$-bins of width 0.2 and that we, therefore, do not extend $\cos(\theta)$-axis beyond the bin centres at -0.9 and 0.9. We see a clear apparent overdensity at low redshifts and an apparent underdensity at high redshifts in the direction that we are headed, i.e. $\cos(\theta) = 1$. This can be understood as the rocket effect blueshifts all objects where the line of sight coincides with our local motion. Thus, at redshifts below (above) the peak in $\bar N(z)$, objects appear at a redshift with lower (higher) $\hat N(z, \theta)$, causing the apparent over-(under)density at redshifts below (above) the peak of $\bar N(z)$.}
    \label{fig:delta_z_theta}
\end{figure}

In Figure \ref{fig:delta_z_theta}, we show the density field $\delta_{\rm rocket}$ as a function of redshift and angle $\theta$;  the black contours are predicted from eq. \eqref{eq:delta_rocket_z} while the colours show the signal estimated from the mock realization.  We bin the number of random objects $R(z, \theta)$ and that of shifted objects $D(z, \theta)$ in redshift $z$ and angular separation to the dipole $\theta$. Figure \ref{fig:delta_z_theta} shows  that the $D(z,\theta)/R(z,\theta)-1$ estimated from our mock realisation faithfully reproduces eq. \eqref{eq:delta_rocket_z}.

\begin{figure}
    \centering
    \includegraphics[width =0.8 \textwidth]{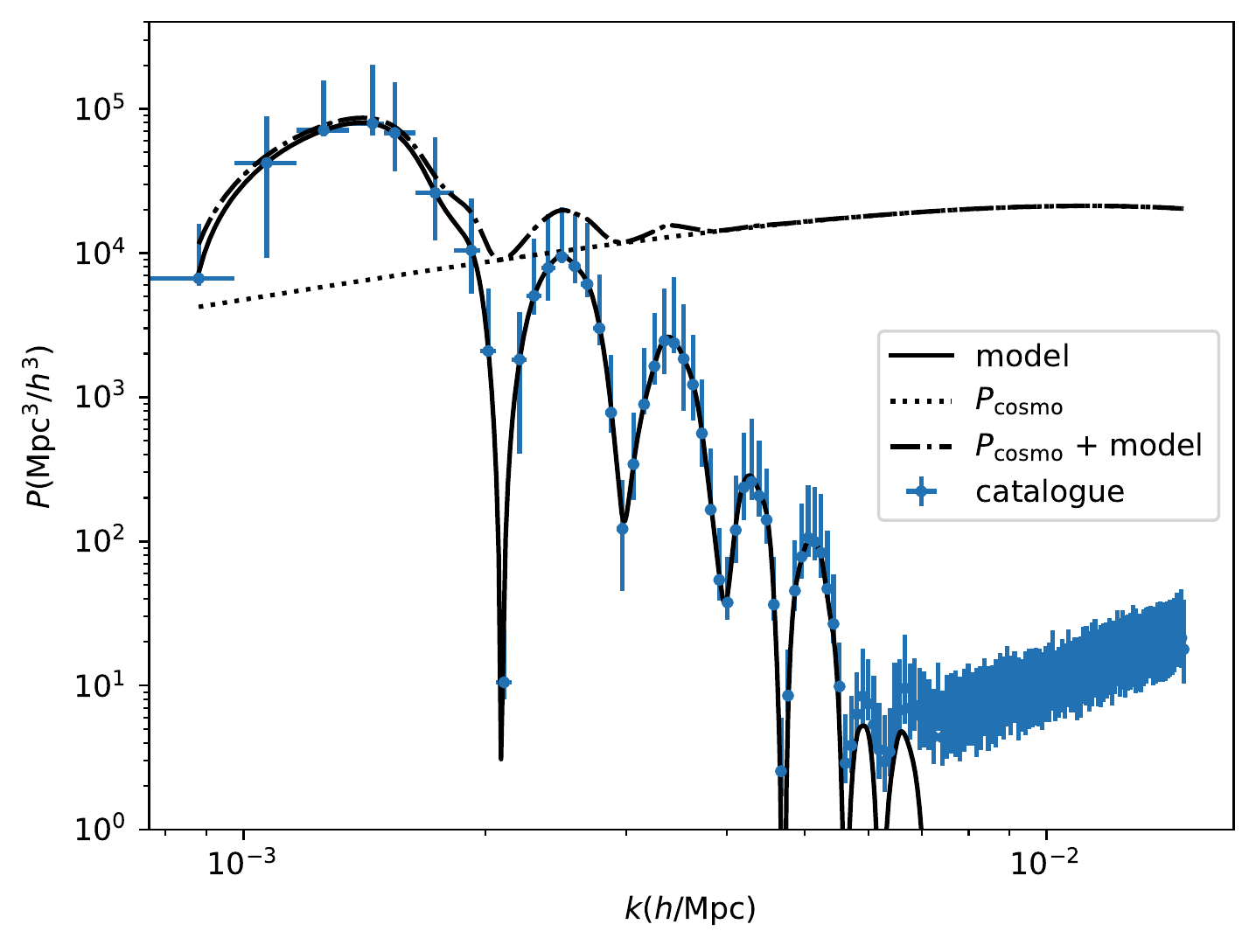}
    \caption{Comparison of the model rocket power spectrum for a Gaussian radial selection function $N(z)$ (eq. \eqref{eq:Gaussian_n_of_z}) with the rocket power computed from a mock realisation of a full-sky survey with the same $N(z)$ in log-scale. The excess power measured from the mock realisation seen at large $k$ is due to discreteness and the mass assignment scheme (see text for more details), in a practical application it is however completely negligible. To better illustrate this fact, we also plot a cosmological model power 
     spectrum and the sum of the cosmological and model Kaiser rocket spectrum as a dot-dashed line. The oscillations are induced by the $j_1$ in eq. \eqref{eq:rocketFB2}.}
    \label{fig:Gaussian_P}
\end{figure}

\begin{figure}
    \centering
    \includegraphics{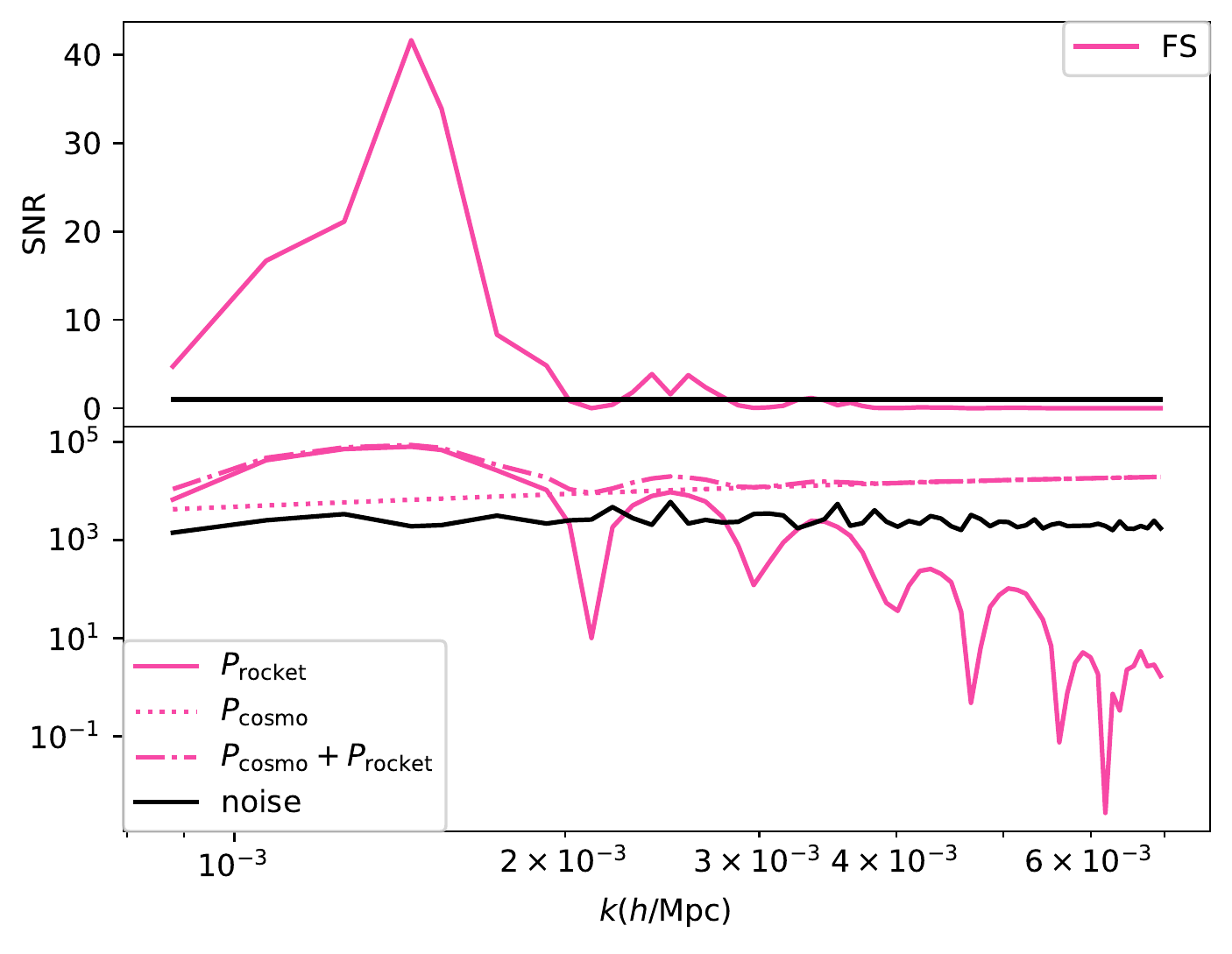}
    \caption{Magnitude of the Kaiser Rocket effect on the power spectrum for a full-sky survey with Gaussian radial selection function (Toy Model I, cf. \ref{sec:Gaussian_toy}). Top panel: Systematic (Rocket) to noise ratio. Bottom panel: Kaiser rocket power spectrum and corresponding noise (which is heavily dominated by cosmic variance). The black line guides the eye to where the SNR 
    per $k$-bin is unity. We show the rocket power spectrum as a solid line, the cosmological model power 
    spectrum used when computing the noise as a dashed line, and the sum of the two as a dot-dashed line.}
    \label{fig:Gaussian_SNR}
\end{figure}

As a next step, we estimate $\delta_{\rm rocket}(\mathbf{k})$ using the FKP estimator on a $512^3$-grid with a side length of 12 Gpc/$h$ to then obtain a power spectrum measurement. Light-cone effects are ignored. 
We obtain the error bars presented in Figure \ref{fig:Gaussian_P} by counting the 1-$\sigma$ percentiles of the modes in each bin. We observe that the median power spectrum is indeed following the model rocket power. The model rocket power (black line in the figure) is obtained from eq. \eqref{eq:Pk_rocket}. Both power spectra are oscillating with increasing amplitude as we go to larger scales. This oscillating can be understood in the limit where the Gaussian selection function is infinitely narrow. In this limit, $\mathcal{N}$ is proportional to a Dirac delta distribution, reducing the scale dependence of eq. \eqref{eq:rocketFB2} to the first spherical Bessel function $j_1$.  Only at small scales, we can see in the log-log plot a noise bias; we have verified that it is due to aliasing and it would disappear by increasing the grid volume, the number of random objects and by using a more sophisticated mass assignment scheme, e.g.,  \cite{Cui:2008fi, Sefusatti:2015aex}. More pragmatically, in the presence of (cosmological) clustering signal, this noise bias will be subdominant and can be ignored for the purpose of this work, as on those scales ($k>7\times 10^{-3}$ h/Mpc) the rocket effect signal effectively vanishes.  We show this in Figure \ref{fig:Gaussian_SNR} where we plot the systematic to noise ratio (SNR, we do not call it signal-to-noise ratio to emphasise the fact that this is not a cosmologically interesting signal but a potential systematic effect on the measured power spectrum) of the Kaiser rocket signal $P_\mathrm{rocket}(k)$ with respect to the noise in the presence of a cosmological signal obtained assuming Gaussian statistics. This is given by  
\begin{equation}
N(k) = \sqrt{\frac{2}{M(k)}}\left[P_\mathrm{cosmo}(k) + P_\mathrm{Al}(k)\right]
\end{equation}
where $M(k)$ denotes the number of modes, $P_\mathrm{cosmo}(k)$ is the cosmological power spectrum that we computed using the Cosmic Linear Anisotropy Solving System  \cite[CLASS]{Blas:2011rf,DiDio_Classgal} for its default cosmology, and we model the noise bias seen in Figure ~\ref{fig:Gaussian_P} as a simple power law $P_\mathrm{Al}(k) = A k^{B}$ 
when computing the SNR from the mock realisation and as $P_\mathrm{Al}(k) = 0$ in the theory case. We obtain the parameters $A = 4563$ and $B = 1.9$ of $P_\mathrm{Al}$ by fitting a power law to the rocket power spectrum at scales $k>7\times 10^{-3}$ h/Mpc. We acknowledge that this SNR estimation is simplistic,  but serves to quantify the scales that are affected by the Kaiser rocket effect. 

\subsection{The Scaling of the Rocket Power Spectrum with the Dipole Amplitude}

Next we check whether the scaling of the rocket power spectrum (cf. eq. \eqref{eq:deltarocketz} and \eqref{eq:Pk_rocket}) with the dipole amplitude (expressed as the velocity $v$) holds as well in our mock realisations. As we know from eq. \eqref{eq:deltarocketz} that $\delta(\mathbf{k})\propto v$, we expect the rocket power spectrum $P(k)\propto v^2$. We repeat the previous analysis for a dipole with twice and four times the Planck velocity. Even though Planck measured the amplitude of the CMB dipole with sub-per mille precision, we do not know our local motion with the same precision considering that the CMB dipole might have other components such as a primordial dipole or contributions from local structure through the integrated Sachs-Wolfe effect. A Bayesian estimate \cite{Saha:2021bay} using higher multipoles of the Planck-2018 CMB temperature map places the maximum posterior value of our local velocity at $v = 0.996 \times 10^{-3}$ with its 68 per cent credible interval extending between $0.83<v\times 10^3 < 1.27$. Furthermore, at lower redshifts, our bulk motion dipole can have additional components. Measurements of our motion dipole from the NRAO VLA Sky Survey (NVSS) suggest an up to four times larger dipole velocity \cite{Blake:2002gx,Singal:2011dy,Gibelyou:2012ri,Rubart:2013tx,Kothari:2013gya,Tiwari:2013ima,Tiwari:2015tba,Siewert:2020krp}, with  direction consistent with the CMB one. A good understanding of the velocity scaling of the rocket power spectrum is therefore important when estimating how the rocket power spectrum biases cosmological measurements.
Our numerical implementation confirms that at ultra-large scales the rocket power we implement on our mocks indeed scales with $v^2$. There are only deviations from the $v^2$-scaling behaviour where the amplitude of the rocket power is small and where the SNR goes to zero so this deviation is not significant. 

\subsection{Toy Model II: Euclid-like Selection Function}
\label{sec:Euclid_toy}

As a first step towards forecasting how future surveys are affected by the Kaiser rocket effect, we increase the realism of the mocks  by choosing a selection function given by
\begin{equation}
       \bar N(z) = 1815\exp\left(-\frac{1}{2}\left[\frac{z-0.99}{0.57}\right]^2\right).
    \label{eq:dNdOmdz_model}
\end{equation}
Our choice is consistent with the $\bar N(z)$ given in \cite{Blanchard:2019oqi} and our results are thus suitable to estimate the impact of the Kaiser rocket effect on a Euclid-like survey.
We generate random objects throughout the redshift range $0.1 < z < 2.1$. Thus, we ensure that edge effects are properly accounted for in generating the mocks. In fact, objects that in the absence of the rocket effect would not be observed can appear in the targeted volume. However, when computing the rocket power spectrum, we only consider objects within the range $0.9 < z < 1.8$. 

For a fair comparison with the theory rocket power spectrum, we only perform the radial Fourier integral in eq. \eqref{eq:deltarocketz} over the redshift range of our Euclid-like mock survey. Alternatively, we could also cut $\bar N(z)$ in a top-hat manner at the redshift boundaries of the survey, which would result in a Dirac delta in the log-derivative. Doing so would be straightforward in the analytic calculations but would not be well-defined in the mocks.

\begin{figure}
    \centering
    \includegraphics{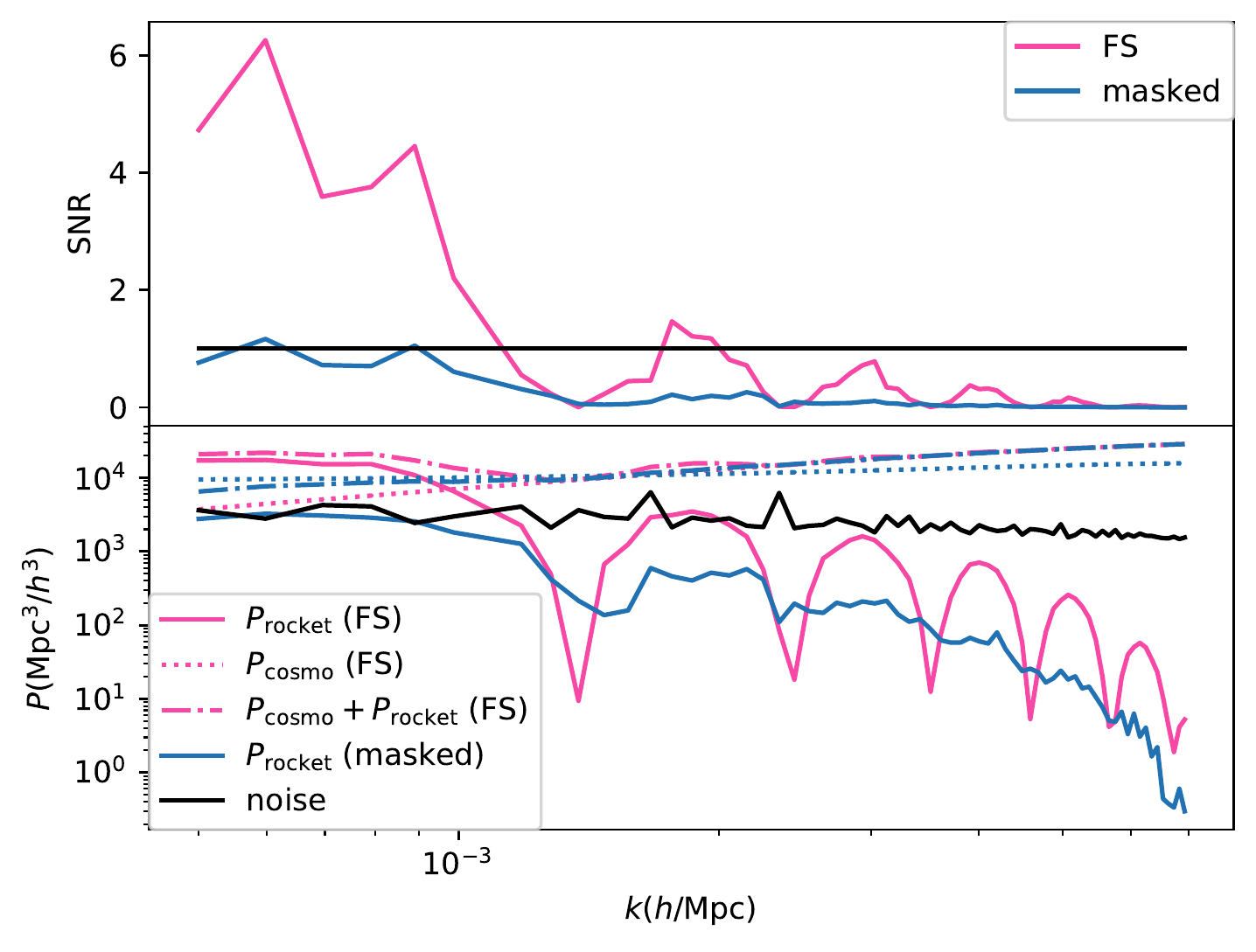}
    \caption{SNR-plot similar to Figure \ref{fig:Gaussian_SNR} but for the Euclid-like full-sky toy model of Section \ref{sec:Euclid_toy}. Even though our Euclid-like selection function is close to Gaussian, differences arise because the peak is at redshift $z\sim 1$ compared to $z = 1.5$ in Figure \ref{fig:Gaussian_SNR}, and because there is a redshift cut at $z = 0.9$, thus omitting most of the left slope of the selection function. Furthermore, the shot noise is estimated with a Euclid-like survey in mind. We show the SNR and power spectra estimated from random catalogues covering both the full (FS) and cut sky (masked). We can see that the mask induced mode coupling washes out the oscillatory features of the full-sky rocket power spectrum. The cut-sky signal is further reduced because much of the area around the motion dipole lies outside the survey footprint. As in Figure \ref{fig:Gaussian_SNR}, we show the rocket power spectrum as a solid line, the cosmological model power 
    spectrum used when computing the noise as a dashed line, and the sum of the two as a dot-dashed line. The masked $P_\mathrm{cosmo}$ is convolved with the window function as decribed in Section \ref{sec:survey_mocks}.}
    \label{fig:Euclid_SNR}
    \label{fig:P_rocket_Euclid_FS_vs_masked}
\end{figure}

The model and mock rocket power spectra for this Euclid-like full-sky configuration are plotted in the lower panel of Figure \ref{fig:Euclid_SNR}. The Kaiser rocket effect overall causes less spurious power than in toy model I Section~\ref{sec:Gaussian_toy}, Figure ~\ref{fig:Gaussian_P}. This decrease is predominantly due to the redshift cut. As the minimum redshift is close to the peak in $\bar N(z)$, this set-up does not cover the redshift range with a sizeable $\bar N(z)$-gradient at redshifts below the $\bar N(z)$ peak. The power spectrum would be closer to the one we obtained in toy model I  if we would also cover redshifts down to $z\sim 0.5$. As the overall systematic signal is lower than in the full Gaussian case, there is less signal bleeding into smaller scales, and we can notice the standard white shot noise with the noise bias suppressed.

We perform a similar  SNR estimation as for the previous toy model, which we plot in Figure \ref{fig:Euclid_SNR}.  The Kaiser rocket effect is subdominant compared to the noise at all scales. However, for a realistic estimate, we also have to consider the effect of the survey mask.

\subsection{Mocks with Realistic Survey Windows}
\label{sec:survey_mocks}

Having seen that the signal induced by the Kaiser rocket effect can be reproduced well using mock realisations with a given radial selection function,  we can proceed to consider more realistic survey configurations which make the rocket signal not analytic and study the effect exclusively  with mock realisations.

\begin{figure}
    \includegraphics[width=\textwidth]{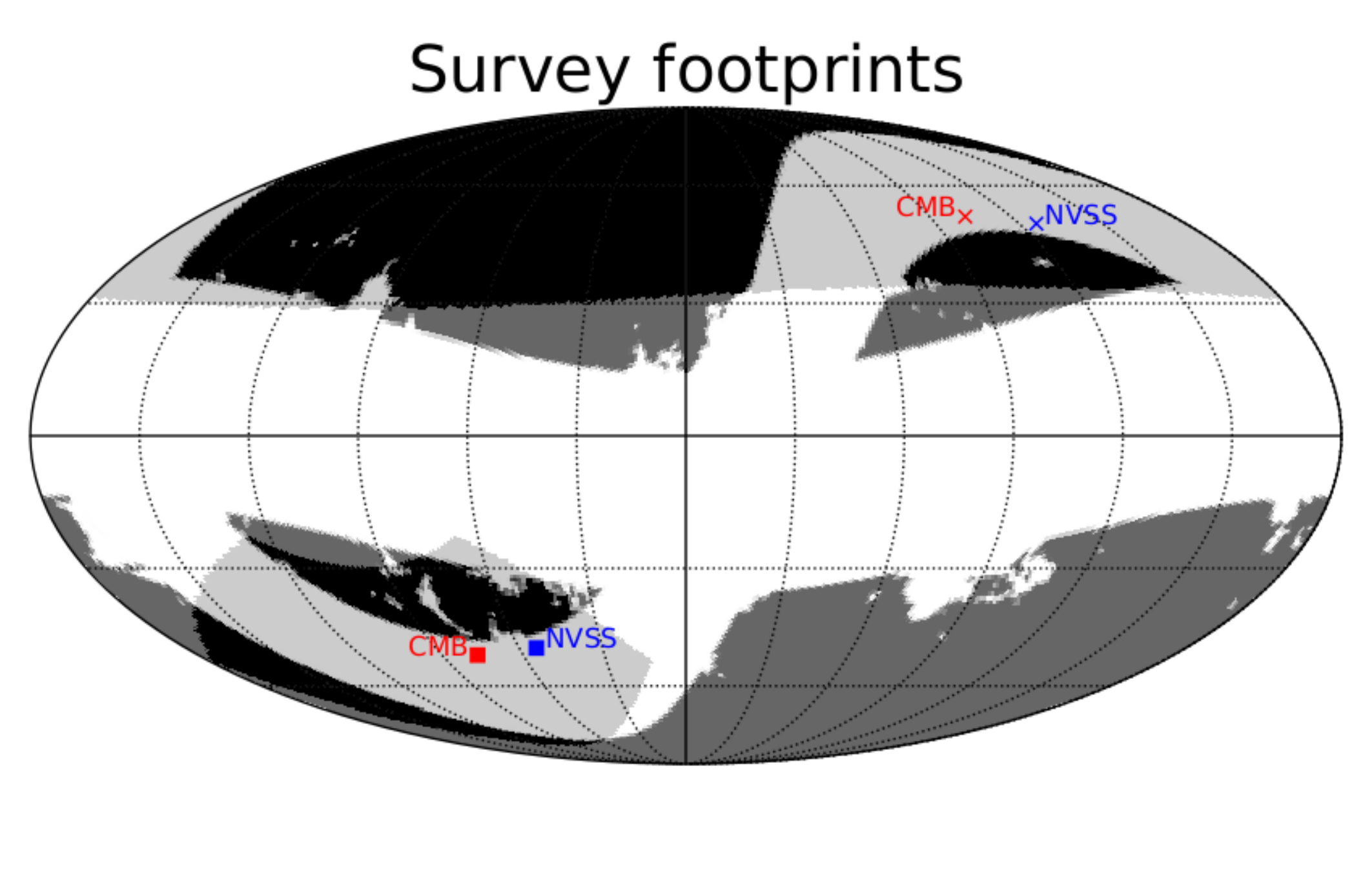}
    \caption{Footprints considered in this work along with the CMB and NVSS dipole directions (marked with crosses) and their respective anti-poles (marked with squares). We colour the Galactic coordinates covered by DESI-like masks in light grey, by Euclid-like masks in dark grey, and areas that are covered by both in black in Mollweide projection. The map is arranged such that the Galactic East is on the left-hand side.}\label{fig:survey_footprints}
\end{figure}

 As modelling the full window function of a given survey is most natural in a random catalogue, we introduce an additional step in the generation of our mock catalogues described above. For instance, to apply the survey mask to a Euclid-like random, while generating the random catalogue we discard proposed  positions outside the Euclid-like footprint,\footnote{We obtained  the footprint as a mangle file 
\url{http://www.mpe.mpg.de/~tdwelly/erosita/multiwavelength_coverage/mangle_polygons/Euclid/Euclid_wide_survey_CORE.ply}  following Ref.~\cite
{Swanson:2007aj}.} which is illustrated in Figure  \ref{fig:survey_footprints} along with the dipole direction from Planck and NVSS.

We are aware that in a real survey the window function might not be separable in angular and radial part as we have assumed here for simplicity. This approximation should however suffice here to quantify the effect and its possible impact on cosmological inference. 
\begin{figure}
    \includegraphics[width=\textwidth]{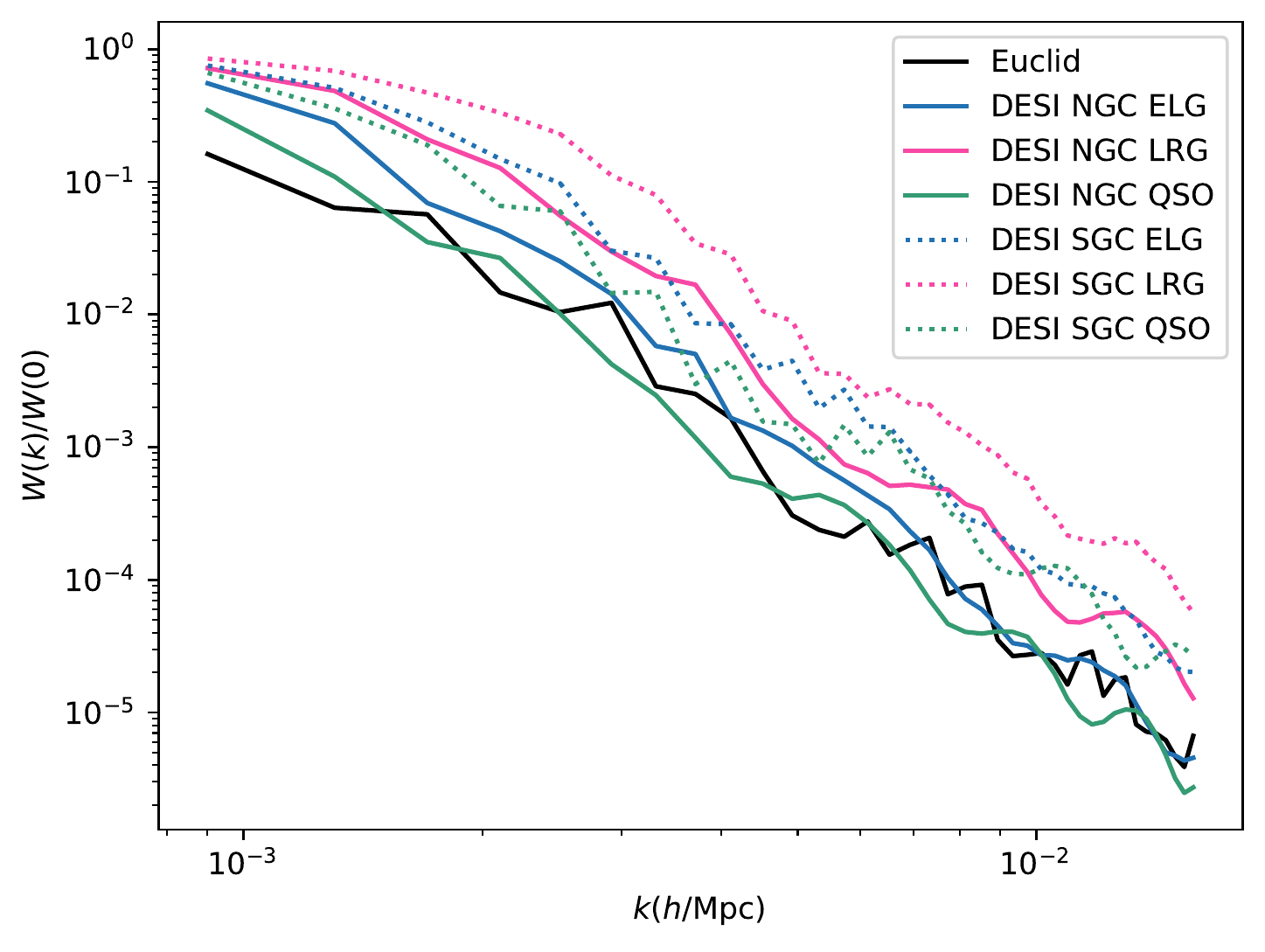}
    \caption{The modulus of the Fourier transform of the survey window function $W(k)$ estimated from the masked random catalogues.} 
    \label{fig:Euclid_masked_lwin}
\end{figure}

We expect the survey window to impact the rocket power spectrum in two ways: {\it i)}
 it  induces mode coupling in Fourier space, between small- (less affected)  and large-scale (more affected) modes, and 
    {\it ii)} it breaks the spherical symmetry of the full-sky configuration and, as a consequence, the position of the dipole (with respect to the mask) matters.

To estimate the effect of mode coupling, we compute the power spectrum of the window $W(k)$ as the band power average (averaging over direction) over a shell in k-space of
\begin{equation}
 \frac{W({k})}{W(0)}=\frac{\left\vert\int\d r\; r^2 \bar n(r)w(r)\frac{\sin(kr)}{kr}\right\vert^2}{\left\vert\int\d r\; r^2\bar n({ r})w(r)\right\vert^2}  
 \label{eq:windowpower}
\end{equation}
from the masked mock catalogue using an FKP estimator  \cite{Feldman:1993ky}, where $\bar n(r)$ is the angle averaged number density in a shell at distance $s$. Following FKP,  $w$ in this equation  denotes the weight which we take to be $$w(r)=\frac{1}{1+\bar n(r)P(k)}\,.$$We present the result in Figure \ref{fig:Euclid_masked_lwin}. The power spectrum of the masked catalogue is given by the convolution of $W(k)$ and the full-sky power spectrum.  We compare the full-sky and masked power spectra in Figure \ref{fig:P_rocket_Euclid_FS_vs_masked} where one can appreciate that mode coupling indeed washes out the oscillatory features in the rocket power spectrum.  This makes it more similar in shape to the power spectrum boost we would expect due to a positive value of the primordial non-Gaussianity parameter $f_\mathrm{NL}^\mathrm{(loc)}$. We shall study the effect this degeneracy has on inferring $f_\mathrm{NL}^\mathrm{(loc)}$ in Section~\ref{sec:fisherbias}.  

Since the rocket effect vanishes on the plane orthogonal to the velocity vector and is maximal in the direction parallel to it, a survey whose footprint is mostly aligned with (orthogonal to) the dipole is maximally (minimally) affected by the Rocket signal. However surveys, in general, avoid the plane of the galaxy, some surveys avoid also the ecliptic, while maximising the fraction of the sky observed,  leaving relatively little freedom to select the orientation of the footprint with respect to the dipole.
To illustrate the importance of the orientation of the dipole with respect to the survey area, we consider two different dipole configurations that we also set in position with respect to the survey footprints in Figure \ref{fig:survey_footprints}:
\begin{itemize}
    \item Assuming an isotropic and homogeneous universe, the dipole signal that is prominent in the cosmic microwave background (CMB) is interpreted as being due to our motion with respect to the CMB rest frame. Recent results by Planck \cite{Akrami:2018vks} quote a dipole signal that corresponds to a velocity of $v=(1.23357 \pm 0.00036) \times 10^{-3}$ (in natural units) in a direction with right ascension (RA) $167.^\circ942 \pm 0.^\circ 007$ and declination (DEC) $-6.^\circ 944 \pm 0.^\circ 007$ (J2000). We expect the LSS dipole to be dominated by the CMB dipole. Not assuming a purely kinematic interpretation of the CMB dipole, \cite{Saha:2021bay} report our local velocity as $v = 0.996^{+0.27}_{-0.17} \times 10^{-3}$ and the direction of our local motion as $(\ell, b) = (268.5^\circ \pm 49.8^\circ, 61.8^\circ \pm 12.3^\circ)$.
    \item Interestingly, the NRAO VLA Sky Survey (NVSS) supports the CMB dipole direction. However, its strong dipole amplitude is still puzzling \cite{Blake:2002gx,Singal:2011dy,Gibelyou:2012ri,Rubart:2013tx,Kothari:2013gya,Tiwari:2013ima,Tiwari:2015tba}. As we expect the LSS dipole to be dominated by the CMB dipole, we consider the NVSS dipole as a "worst-case" deviation from the CMB dipole. For our analyses, we adopt $\mathrm{(RA,DEC)}=(154,-2)^\circ\pm 19^\circ$ and $v=(4.5\pm 1.5)\times 10^{-3}$ from \cite{Rubart:2013tx}.
\end{itemize}
We have already established that the rocket power spectrum scales well with $v^2$ at large scales. What remains to study is the effect of varying the direction of the dipole. We show this effect summarised as a shift of the best fitting parameter value of $f_\mathrm{NL}$ in Figure \ref{fig:DeltafNL_map}.

\begin{figure}
    \centering
    \includegraphics[width=\textwidth]{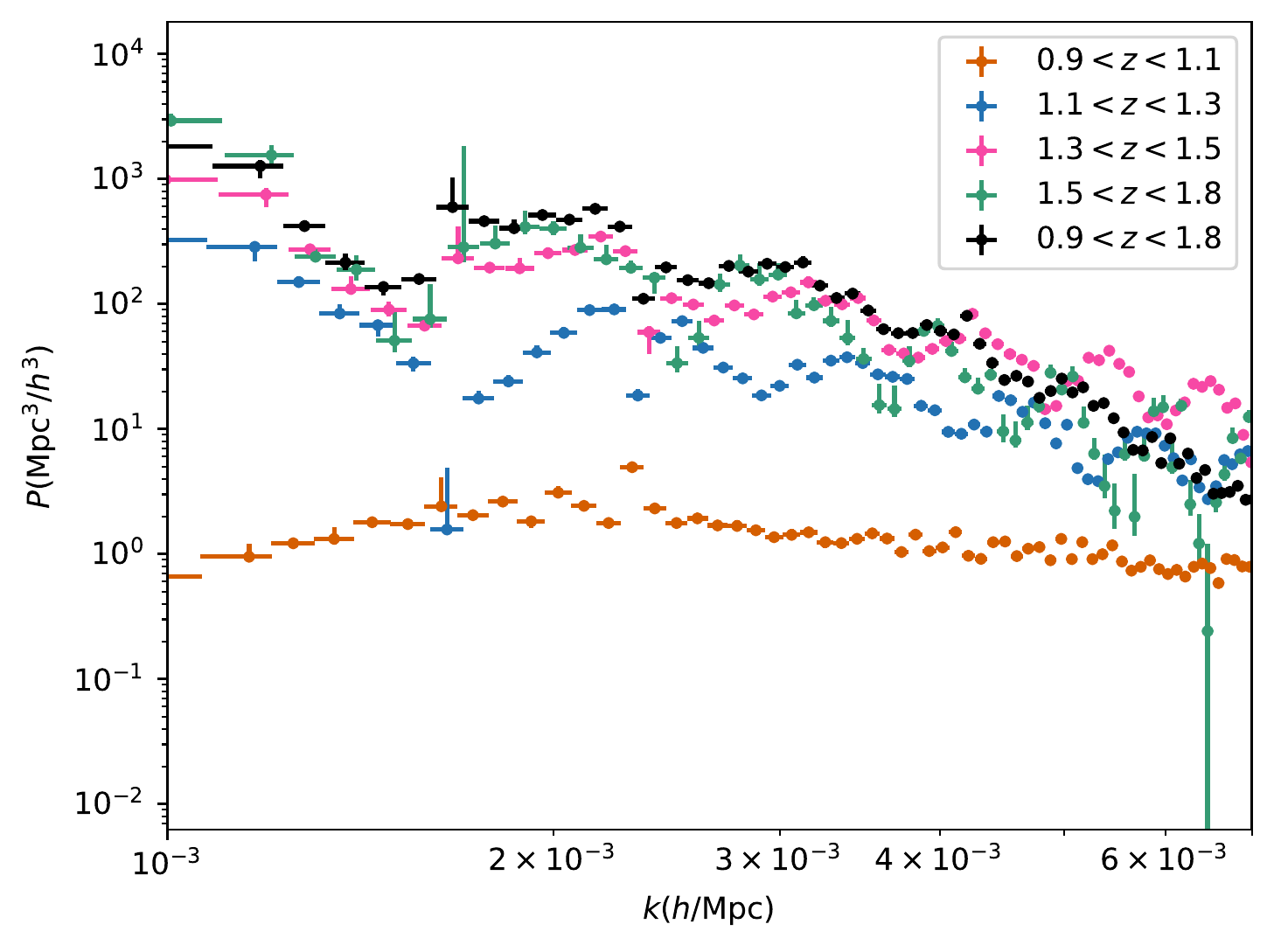}
    \caption{The power spectrum induced to a Euclid-like mock catalogue by the Planck-CMB dipole. The black points have been obtained from the whole sample, while for the other ones, we have split our mock catalogue into four redshift slices. The data points of the power spectra in redshift bins have been shifted slightly along the $k$-axis for clarity and they correspond to the closest data point in black. We cut the power spectra at $k = 0.007\;h/\mathrm{Mpc}$ where the noise dominates the systematic signal.}
    \label{fig:P_rocket_binned}
\end{figure}

For some physical processes affecting large scales it is beneficial to use the complete available survey volume (e.g.,  to constrain large-scale parameters such as $f_\mathrm{NL}^\mathrm{(loc)}$) using redshift weights \cite{Zhu:2014ica, Ruggeri:2016mac,Mueller:2017pop,Ruggeri:2017rza,Ruggeri:2020cus}, but the standard approach is to split the survey into redshift shells to account for the evolution of the background cosmology. Therefore, we also consider the latter approach and, inspired by \cite{Blanchard:2019oqi}, split our Euclid-like mock realisation into four redshift bins with boundaries at $z=\lbrace 0.9, 1.1, 1.3, 1.5, 1.8\rbrace$. We display the rocket power spectra of each redshift slice along with the full-survey one in Figure \ref{fig:P_rocket_binned}. The power spectra above $z>1.1$ follow similar trends, while the lowest redshift bin at $0.9<z<1.1$ does not reveal any sizeable rocket power. This lack of power can be understood considering that this redshift bin covers the region around the peak of $\bar N(z)$ where the average number density distribution $\bar N(z)$ is locally flat; its log-derivative is therefore negligible, so is the rocket power spectrum (cf. eq. \eqref{eq:delta_rocket_z}). As the redshift increases in the other bins, $\bar N(z)$ becomes steeper, and we can also perceive an increase in the rocket power in Figure \ref{fig:P_rocket_binned}.

\section{The Bias on Primordial Non-Gaussianity Measurements}
\label{sec:fisherbias}
Not accounting for the Kaiser rocket effect adds power at ultra-large scales (see also \cite{Bertacca:2019wyg}). This systematic increase in power has the potential to systematically bias the best-fitting parameters that we obtain from the incorrectly estimated power spectra.
In particular, a natural worry then is that the extra power induced by the Kaiser rocket effect, if unaccounted for, could mimic the signal of a small primordial non-Gaussianity of the local type. Here we  estimate the bias on recovered cosmological parameters of a $f_{\rm NL} \Lambda $CDM-model -- a  $\Lambda $CDM-model with an extra parameter  for the amplitude of a (small) primordial local non-Gaussianity-- in the presence of unsubtracted Kaiser rocket effect. We adopt the forecasting approach  described in \cite{Taylor:2006aw,Heavens:2007ka,Taruya:2010mx,Taylor:2012kz,Duncan:2013haa,Natarajan:2014xba,Camera:2014sba,Pullen:2015yba,Cardona:2016qxn,Sellentin:2017fbg,Raccanelli:2017kht,Kodwani:2018uaf,Mike:2018zvb,Jelic-Cizmek:2020pkh,Bernal:2020pwq}.
\subsection{Systematic Shift of the Best-Fit Parameters from incorrect rocket modelling}
 For a Gaussian-distributed observable $\Psi$, one can estimate the systematic shift $\triangle_\mathrm{syst} = \triangle\vartheta^\mathrm{(I)} - \triangle\vartheta^\mathrm{(C)}$ of the best-fitting parameters $\vartheta^\mathrm{bf, I}$ and $\vartheta^\mathrm{bf, C}$ from a measurement where $\Psi$ is modelled incorrectly or correctly, respectively, using \cite{Taylor:2006aw,Heavens:2007ka,Taruya:2010mx,Taylor:2012kz,Duncan:2013haa,Natarajan:2014xba,Camera:2014sba,Pullen:2015yba,Cardona:2016qxn,Sellentin:2017fbg,Raccanelli:2017kht,Kodwani:2018uaf,Mike:2018zvb,Jelic-Cizmek:2020pkh,Bernal:2020pwq}
 \begin{equation}
    \triangle_\mathrm{syst} = F_\mathrm{I}^{-1}\sum_{ij}\left(\nabla_\vartheta \hat\Psi_i^\mathrm{(fid, I)}\right)\left(C^{-1}\right)_{ij}\left(\hat\Psi_j^\mathrm{(fid, I)}-\hat\Psi_j^\mathrm{(fid, C)}\right),
    \label{eq:Fisher_shift}
\end{equation}
where $F_\mathrm{I}$
is the Fisher information matrix of the incorrect model and $\nabla_\vartheta$ represents the gradient in parameter space.

Before applying eq. \eqref{eq:Fisher_shift} to estimate the rocket-induced shift, 
we have to pick an observable. So far, we have discussed the effect of the Kaiser rocket effect on the galaxy power spectrum; at the ultra-large scales where the Kaiser rocket effect has an impact, the number of Fourier modes is low and the assumption of Gaussianity, underlying this section, is not valid. Nevertheless, this can be saved by improving data normality by carrying out a Box-Cox transformation \cite{BoxCox,Joachimi:2011iq,Schuhmann:2015dma,Mike:2018zvb}. Alternatively, for the purpose of forecasting, assuming an approximately Gaussian over-density field $\delta(\mathbf{k})$, the observable 
\begin{equation}
    \Psi = P_\mathrm{fid}(k)\sqrt{\frac{P_\mathrm{d}(k)}{P(k, \vartheta)}-\ln\left(\frac{P_\mathrm{d}(k)}{P(k, \vartheta)}\right)}
\end{equation}
is also approximately Gaussian \cite*{Hamimeche:2008ai,Kalus:2015lna}, where $P_\mathrm{d}(k)$ is the data power spectrum and $P_\mathrm{fid}(k)$ the fiducial one, and all power spectra are sufficiently close to the true power spectrum. Note that $P_\mathrm{fid}(k)$ is a fixed power spectrum for an arbitrary set of parameters that correspond to the ones used to compute the covariance matrix, and it is not necessarily equal to $P_\mathrm{fid, C}(k_i)$ or $P_\mathrm{fid, I}(k_i)$, even though it makes sense to make this choice. It is, however, important to keep that in mind when taking derivatives. With the  assumption that $\Psi^\mathrm{(d)} = \hat\Psi^\mathrm{(fid, C)}$, the ingredients to eq. \eqref{eq:Fisher_shift} thus read
\begin{align}
    \Psi^\mathrm{(fid, C)}_i &= P_\mathrm{fid, C}(k_i),\nonumber\\
    \Psi^\mathrm{(fid, I)}_i &= P_\mathrm{fid, I}(k_i)\sqrt{\frac{P_\mathrm{fid, C}(k_i)}{P_\mathrm{fid, I}(k_i)}-\ln\left(\frac{P_\mathrm{fid, C}(k_i)}{P_\mathrm{fid, I}(k_i)}\right)},\nonumber\\
    \nabla_\theta \hat\Psi_i^\mathrm{(fid, I)} &= \frac{P_\mathrm{fid, I}(k_i) - \frac{P_\mathrm{fid, C}(k_i)}{P_\mathrm{fid, I}(k_i)}}{2\Psi^\mathrm{(fid, I)}_i}\nabla_\theta P_\mathrm{fid, I}(k_i).
    \label{eq:GaussianisedPower}
\end{align}
    
We use CLASS \cite{Blas:2011rf,DiDio_Classgal} to compute model power spectra $P_\mathrm{fid, C}(k)$. To obtain $P_\mathrm{fid, I}(k)$, we add the rocket power spectrum estimated from shifted random catalogues (cf. Section \ref{sec:meth}) to the correct power spectrum $P_\mathrm{fid, C}(k)$. The window power $W(k)$ is also computed from the random catalogue using eq. \eqref{eq:windowpower}.

\subsection{The Full-Sky Euclid-like Case}
At large scales, where the Kaiser rocket effect adds power if not accounted for, local primordial non-Gaussianity can also boost power as it alters in a scale-dependent way  the biasing law between dark-matter halos and the underlying mass-density field \cite{Dalal:2007cu, Matarrese:2008nc, Slosar:2008hx, Afshordi:2008ru, Valageas:2009vn, Giannantonio:2009ak, Schmidt:2010gw, Desjacques:2011jb}.
At the ultra-large scales that are of interest, 
the galaxy power spectrum monopole can be described by the linear Kaiser redshift space distortion model \cite{Kaiser:1987qv}
\begin{equation}
    P^{NG}_g(k)=\left(b_\mathrm{NL}^2(k)+\frac{2}{3}b_\mathrm{NL}(k)f+\frac{f^2}{5}\right)P_\mathrm{m}(k),
\end{equation}
where $f$ is the rate of structure growth and $P_\mathrm{m}(k)$ represents the linear matter power spectrum. The coupling of long- and short-wavelength perturbations in the Peak-Background Split model causes the galaxy bias to attain the scale-dependent form
\begin{equation}
    b_\mathrm{NL}(k)=b\left[1+3 \frac{b-p}{b} f_\mathrm{NL}\delta_c(z)\frac{\Omega_\mathrm{m} H_0^2}{k^2}\right]
\end{equation}
where
$b$ is the linear galaxy bias, $p$ is the halo merger bias (which we henceforth set to $p=1$) \cite{Slosar:2008hx}, $\Omega_\mathrm{m}$ is the present day matter density and $\delta_c(z)$ is the critical density in the Peak-Background Split model. We further assume $\delta_c(z)={q \delta_{c,0}/ D(z)}$, with $D(z)$ being the scale-independent growth factor, $\delta_{c,0}=1.686$ the expected critical density from the spherical collapse model in an Einstein-de Sitter
universe, and $q=0.75$ being a fudge factor calibrated to N-body simulations \cite{Wagner:2011wx}.
The measured (total) galaxy  power spectrum is 
\begin{equation}
    \hat{P}_\mathrm{I}(k)=\hat P_\mathrm{C}(k)+P_\mathrm{rocket}(k)
\end{equation}
that is the sum of the "correct" (or cosmological) galaxy power spectrum $P_\mathrm{C}$ and the power spectrum due to the Kaiser rocket effect $P_\mathrm{rocket}$ (see also Section \ref{sec:crosstest}). We use the subscript $I$ on this power spectrum since it is systematically biased (incorrect)  with respect to the cosmological ("correct") one. We assume for forecast purposes that the true power spectrum $$P_\mathrm{true}\equiv P_g^G(k) = \left(b^2 + \frac{2}{3}bf + \frac{f^2}{5}\right) P_\mathrm{m}(k)$$ is given by the  galaxy power spectrum $P_g^G(k)$ of a density field with Gaussian initial conditions. We model the covariance matrix as \cite{Feldman:1993ky,Tegmark:1997rp}
\begin{equation}
    C_{k_1 k_2}=\frac{2}{M}P_g^G(k_1) W^4(k_1-k_2) P_g^G(k_2),
    \label{eq:covariance}
\end{equation}
where 
\begin{equation}
    M=V_n V_\mathrm{eff}(k)
    \label{eq:modes}
\end{equation}
is the number of independent Fourier modes in the $n$th $k$-bin with width $\triangle k_n$ centred around $k_n$,
\begin{equation}
    V_n\equiv \frac{k^2_n\triangle k_n}{2\pi^2}
\end{equation}
is the $k$-space volume of the bin, the effective volume $V_\mathrm{eff}$ is related to the survey volume $V_S$ as
\begin{equation}
    V_\mathrm{eff}(k)\equiv V_S\left[\frac{\bar n P(k)}{1+\bar n P(k)}\right]^2,
\end{equation}
$\bar n$ is the number density averaged over the whole survey volume, and $W(k)$ is the window-induced mode-coupling estimated from the random catalogue (see Section~\ref{sec:survey_mocks}, eq.~\eqref{eq:windowpower} and Ref.~\cite{Feldman:1993ky}).

We start by considering a full-sky analysis case, hence a diagonal covariance matrix where $W(k)=\delta_\mathrm{D}(k)$.  We estimate the systematic shifts in the best-fitting parameters of a $\Lambda$CDM+$f_\mathrm{NL}$ model, that is $\vartheta=\lbrace h, \omega_\mathrm{b}, \omega_\mathrm{cdm}, n_s, b, f_\mathrm{NL}\rbrace$. We choose fiducial values $ h = 0.67556$, $\omega_\mathrm{b} = 0.022032$, $\omega_\mathrm{cdm} = 0.12038$, $n_s = 0.9619$, $b = 1.7$ and $f_\mathrm{NL} = 0$. Ideally, we would also include the amplitude of the power spectrum, as well as $f$ as parameters in the analysis, but they are highly degenerate with the galaxy bias $b$ in the monopole power spectrum for low values of  $f_\mathrm{NL}$. In a more realistic analysis these parameters would be constrained by also considering the power spectrum multipoles; here for simplicity we consider only the power spectrum monopole, hence they are fixed. 

As both the Kaiser rocket effect and local-type primordial non-Gaussianity only affect large scales, one could be tempted to stick to ultra-large scales only. However, at these scales, the parameters of the background cosmology are very poorly constrained, leading to unreasonable shifts especially when assuming an NVSS-like dipole, and through correlations with $f_\mathrm{NL}$, the expected bias on primordial non-Gaussianity constraints are also unreliable. At scales of about $k=0.3\;h/\mathrm{Mpc}$, non-linear effects start to become important at the lower redshift end of the Euclid-like survey.  We obtain very similar estimates for the best-fitting parameter shifts for both $k_\mathrm{max}=0.2\;h/\mathrm{Mpc}$ and $k_\mathrm{max}=0.3\;h/\mathrm{Mpc}$. We, therefore, choose $k_\mathrm{max}=0.2\;h/\mathrm{Mpc}$ for the rest of this work to be slightly conservative, yet having a sufficient $k$-range to anchor the background cosmology. At this point, we just consider the nominal redshift range of Euclid ($0.9<z<1.8$). We also assume here a constant bias $b = 1.7$ taken at the effective redshift $z=1.35$. We account for the redshift evolution of the galaxy bias in the next subsection.

The power spectrum does not obey a normal distribution at scales where the number of modes is small (see discussion in Appendix \ref{app:1}), so we implement the Fisher analysis using the Gaussianised power spectra of eq. \eqref{eq:GaussianisedPower}. The rocket-induced systematic shift in the cosmological parameters is reported in the first two rows of Table \ref{tab:FS_shifts_main} for a full-sky survey with our Euclid-like selection function and for two amplitudes for the dipole (Planck and NVSS).\footnote{In the full sky case the dipole direction has no effect.}. We also plot the shifts in the best-fitting value of $f_\mathrm{NL}$ along with other configurations that we discuss later in Figure \ref{fig:shift_summary}

\begin{table}[ht]
    \centering
    \begin{tabular}{|l|l|c|c|c|c|c|c|}
        \hline & dipole& $\Delta_{\rm syst}$& $\Delta_{\rm syst}$&  $\Delta_{\rm syst}$&  $\Delta_{\rm syst}$&  $\Delta_{\rm syst}$&  $\Delta_{\rm syst}$\\
        & amplitude & $h$ & $\omega_\mathrm{b}$ & $\omega_\mathrm{cdm}$ & $n_s$ & $b$ & $f_\mathrm{NL}$ \\\hline\hline
        Full Sky & Planck  &  2.6$\times 10^{-5}$
 &  -4.5$\times 10^{-6}$
 &  7.1$\times 10^{-5}$
 &  -3.5$\times 10^{-5}$
 &  -0.0010
 &  3.6\\
 & &  0.015 $\sigma$
 &  -0.034 $\sigma$
 &  0.11 $\sigma$
 &  -0.013 $\sigma$
 &  -0.099 $\sigma$
 &  1.1 $\sigma$
\\\hline
Full Sky & NVSS  &  0.00011
 &  -2.1$\times 10^{-5}$
 &  0.00032
 &  -8.9$\times 10^{-5}$
 &  -0.0047
 &  18\\
 & &  0.064 $\sigma$
 &  -0.15 $\sigma$
 &  0.48 $\sigma$
 &  -0.032 $\sigma$
 &  -0.45 $\sigma$
 &  6.2 $\sigma$\\\hline
Masked &        Planck  &  3.7$\times 10^{-5}$
 &  -2.3$\times 10^{-7}$
 &  1.8$\times 10^{-5}$
 &  3.2$\times 10^{-5}$
 &  -0.00054
 &  2.2

\\
& &  0.013 $\sigma$
 &  -0.0016 $\sigma$
 &  0.022 $\sigma$
 &  0.0097 $\sigma$
 &  -0.031 $\sigma$
 &  0.23 $\sigma$
\\\hline
    \end{tabular}
    \caption{Expected shift of the best-fitting $\Lambda$CDM+$f_\mathrm{NL}$ parameters for a full sky/masked survey with our adopted Euclid-like selection function,  and using a Gaussianised version $\Psi(k)$ of the galaxy power spectrum. We use the constraints on the `Base $\Lambda$CDM model' from \textit{Planck} TT,TE,EE+lowE+lensing \cite{Aghanim:2018eyx} as a prior. We assume no prior knowledge on $f_\mathrm{NL}$. We present the shift both as an absolute number as well as in terms of the expected uncertainty of each parameter estimated as $\sigma(\vartheta_\alpha)=\sqrt{(F^{-1})_{\alpha\alpha}}$.}
    \label{tab:FS_shifts_main}
\end{table}
In this case a  Planck dipole amplitude would yield a 1.1$\sigma$ bias on $f_{\rm NL}$, which grows to 6.2 $\sigma$ for an NVSS dipole amplitude. 

In this full-sky scenario, the rocket power spectrum has quite distinctive oscillatory features that make it in principle possible to tell it apart from a power spectrum that is due to primordial non-Gaussianity (cf. Figure \ref{fig:P_rocket_Euclid_FS_vs_masked}). A survey mask washes out these features and, thus, increases the chances of confusing the signals. On the other hand, a survey mask reduces the volume and increases the error bars, which may compensate. 

\subsection{Masked Euclid-like Case}
In reality, any survey will not observe the full sky and, therefore, we have to account for the angular survey window. We do so following \cite{Ross:2012sx}, thus convolving the model power spectrum with a window matrix 
\begin{equation}
    W_{ij} = \int\d\varepsilon\int\d\cos\theta W(\varepsilon)\varepsilon^2 \Theta(r_\varepsilon, k_j ),
\end{equation}
where $r_\varepsilon=\sqrt{k_i^2+\varepsilon^2-2k_i\varepsilon\cos\theta}$, and $\Theta(r_\varepsilon, k_j )$ is one if $r_\varepsilon$ falls into the bin $k_j$ and zero otherwise. The convolved model power spectrum reads
\begin{equation}
    P_\mathrm{conv}(k_i)=\sum_j \left[\left(W_{ij}-W_{0j}\frac{W(k_i)}{W(0)} \right)P_g^{NG}(k_j)\right],
\end{equation}
where the second term in the round brackets is the Fourier space analogue of the integral constraint on the correlation function.

\begin{figure}
    \centering
    \includegraphics[width=\textwidth]{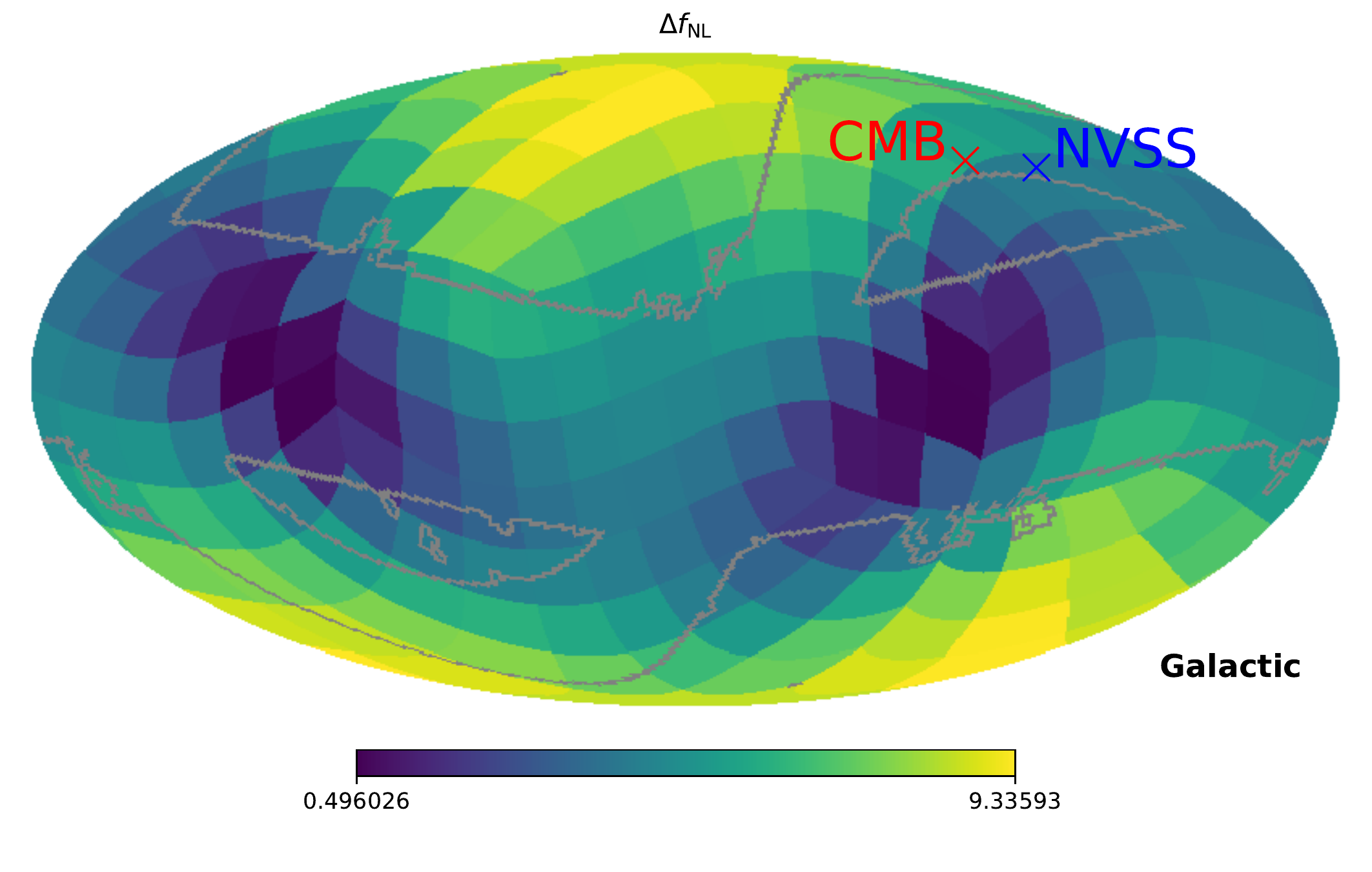}
    \caption{The expected shift in $f_\mathrm{NL}$ due to not accounting for the Kaiser rocket effect under the hypothesis that the observer's motion dipole points towards the centre of each HEALPix cell. The shift is estimated using eq. \eqref{eq:Fisher_shift} assuming a Euclid-like mask and selection function as well as a motion dipole with the amplitude of the CMB dipole measured by Planck \cite{Akrami:2018vks}. For better comparison with Figure \ref{fig:survey_footprints}, we mark the CMB and NVSS dipole directions and plot the contour of the Euclid-like footprint in grey. The map is in Galactic coordinates in Mollweide projection arranged such that the Galactic East is on the left-hand side.}
    \label{fig:DeltafNL_map}
\end{figure}

We start by estimating the shift in best-fit parameters caused by the Planck dipole analogously to the full-sky case. At first, as in the full-sky case, we consider only the redshift range $0.9<z<1.8$. We list the results in the bottom row of Table \ref{tab:FS_shifts_main} and plot them in Figure \ref{fig:shift_summary}, where we  notice that the survey mask slightly decreases the absolute value of the parameter shifts. The shifts become less significant (about a fourth of a $sigma$), as the mask increases the uncertainty on the parameters.  
If the direction of our peculiar motion is very different from the CMB dipole then we can expect the bias on $f_\mathrm{NL}$ measurements to be significantly different. We show in Figure \ref{fig:DeltafNL_map} that if our peculiar motion points close towards the centre of the area covered by the survey, we can expect the effect  to be almost four times as large, thus reaching the 1$\sigma$ level. Even though we do not expect the direction of our peculiar motion to be significantly different from the CMB dipole direction, we have to keep in mind that if inaccurate redshift corrections were to  be applied,  a residual Kaiser rocket signal would remain,  that can have any arbitrary direction.

A redshift dependence of the shifts is expected if the survey were to be split in redshift bins.
In this case, to estimate the parameter biases, we have to take the redshift dependence of the linear galaxy bias into account. Inspired by \cite{Blanchard:2019oqi}, we choose $b_{0.7 < 0.9} = 1.31$, $b_{0.9<z<1.1}=1.46$, $b_{1.1<z<1.3}=1.61$, $b_{1.3<z<1.5}=1.75$ and $b_{1.5<z<1.8}=1.9$. This is another reason why we generated our random catalogues over a wider  redshift range than the fiducial Euclid-like one in Section~\ref{sec:Euclid_toy}.

\begin{figure}
    \centering
    \includegraphics[width = 0.49\textwidth]{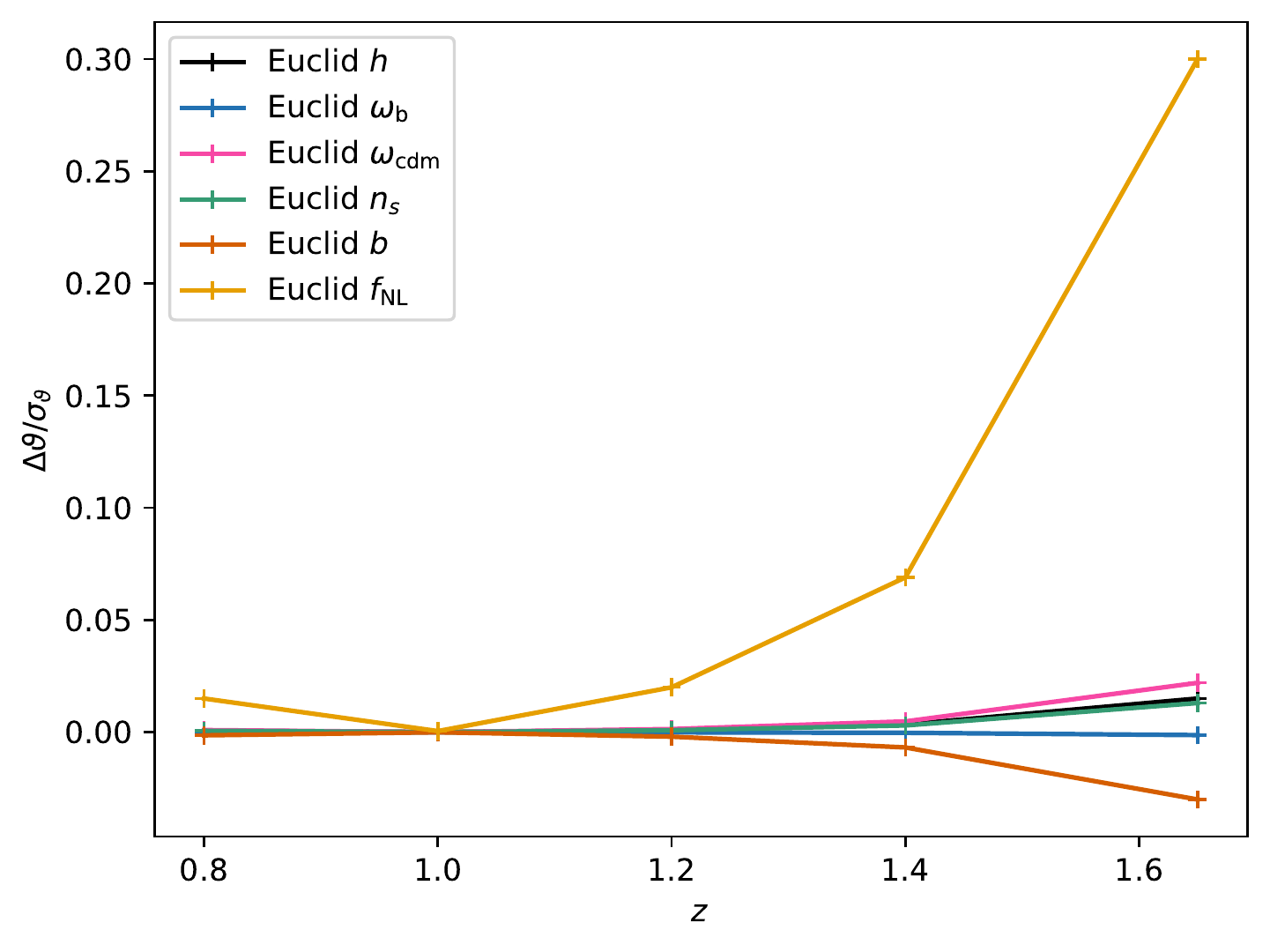}
    \includegraphics[width = 0.49\textwidth]{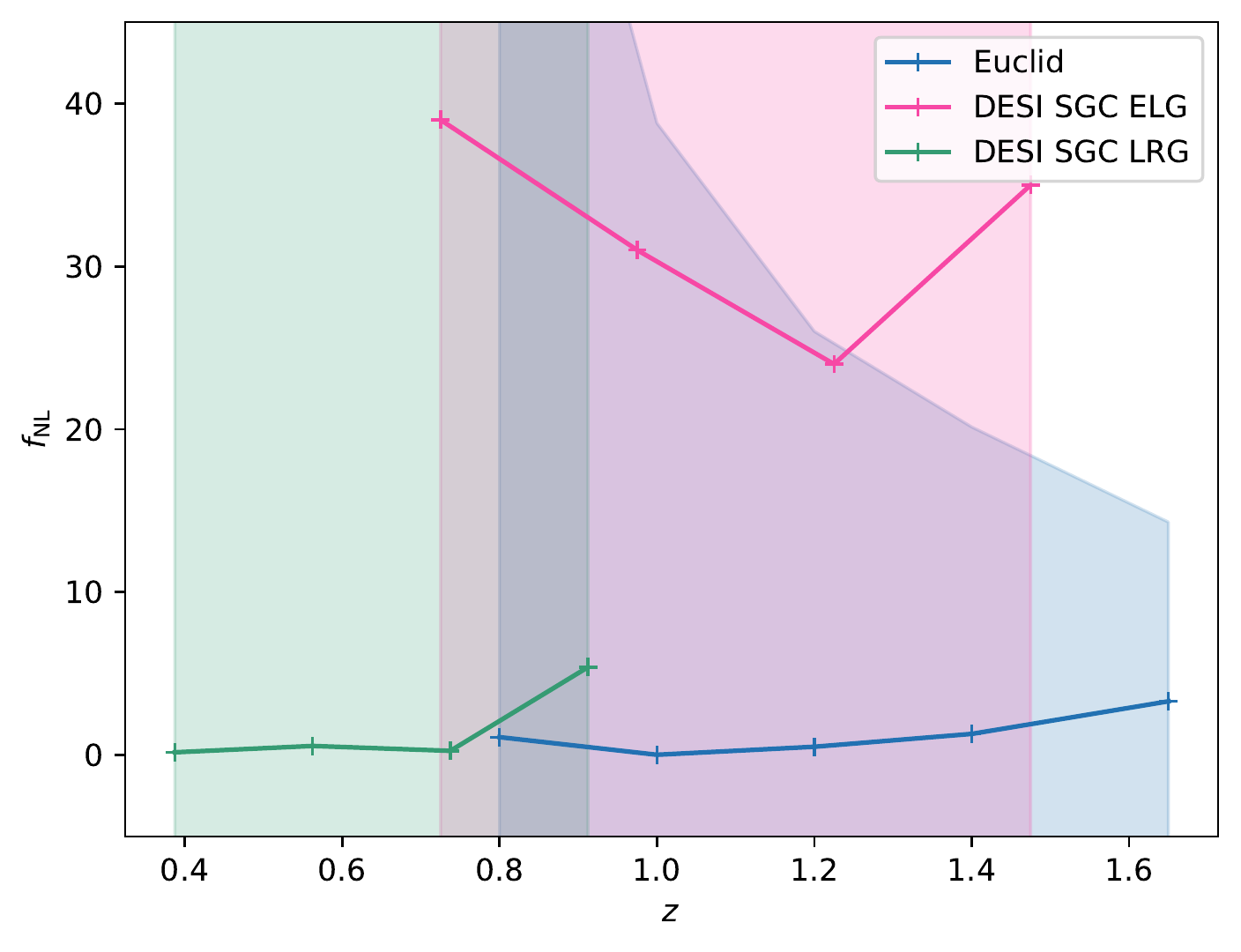}
    
    \caption{Left panel: systematic to statistical error ratio (relative shift) in all parameters for our adopted Euclid-like survey after subdividing the survey into four redshift bins and adding an additional redshift bin below the redshift range covered by the Euclid-like survey. The parameter bias vanishes at the maximum of $\bar N(z)$ which occurs at $z\sim 1$ for Euclid. Right panel: the absolute shift in the best-fit $f_\mathrm{NL}$ value expected from dividing the Euclid-like and DESI ELG and LRG-like surveys into four (five) redshift bins. The DESI forecasts shown here are for the Southern Galactic Cap only, as the effect is largest there. The shaded regions correspond to the 1-$\sigma$ confidence regions forecasted using the same Fisher matrices as those used in predicting the shift in the best-fit. In most cases, these extend over the plotting range and we refer to Figure \ref{fig:shift_summary} for the full error bars. The shift in the best-fitting $f_\mathrm{NL}$ value is expected to be stronger for ELGs than for LRGs due to the fact that ELGs are less biased tracers than LRGs and, therefore, larger values of $f_\mathrm{NL}$ are needed to mimic the Kaiser rocket power spectrum.}
    \label{fig:param_shifts_in_z_bins}
\end{figure}

In Table \ref{tab:FS_shifts_in_z_bins} we list the shifts expected in the four redshift bins listed Section \ref{sec:survey_mocks} and we consequently plot them in Figure \ref{fig:shift_summary}. As we already could expect from the strong redshift dependence of the rocket power spectrum apparent in Figure \ref{fig:P_rocket_binned}, we see only negligible parameter shifts in the first redshift bin at $0.9 < z < 1.1$, where the rocket effect is indeed expected to vanish. To illustrate the redshift dependence of the bias on the estimated parameters at either side of the peak in $\bar N(z)$,  we also consider an additional redshift slice at $0.7 < z < 0.9$. 
  As the derivative of $\bar N(z)$ increases, the parameter shifts also increase. We visualise this dependence in Figure \ref{fig:param_shifts_in_z_bins}. 
  On the left-hand panel, we show the systematic-to-statistical error ratio ($\triangle\vartheta/\sigma_\vartheta$) for the relevant cosmological parameters as a function of the redshift bin. The statistical error is computed from the same Fisher matrix as the best-fit shift. As the shifts are most crucial in $f_\mathrm{NL}$, we report the behaviours of the absolute shift in $f_{\rm NL}$ in the right-hand panel of Figure \ref{fig:param_shifts_in_z_bins}. There, we see a clear redshift dependence of the measured best-fitting $f_\mathrm{NL}$ value stretching from the fiducial value of $f_\mathrm{NL} = 0$ up to $f_\mathrm{NL} = 3.3$. Note that even if $\bar N(z)$ is Gaussian and thus symmetric, the value of $f_\mathrm{NL}$ is not because the galaxy bias approaches unity as we go to smaller redshifts, and thus a higher value of $f_\mathrm{NL}$ is needed to compensate for the rocket power. It is worth noting that another turnover in $f_\mathrm{NL}$ can be expected where the bias parameter goes to values below 1, but this will leave the other parameters unaffected. 

While some variation in the measured value of $f_\mathrm{NL}$ can be expected due to the uncertainty of the redshift evolution of the galaxy bias, to a great degree, finding an extremum in all measured parameters at the redshift where $\bar N(z)$ peaks is a telltale feature of an unaccounted-for velocity dipole in the data.

\begin{table}[ht]
    \centering
    \begin{tabular}{|c|c|c|c|c|c|c|c|}
        \hline $z_\mathrm{min}$ & $z_\mathrm{max}$ & $\Delta_{\rm sys}h$ &
        $\Delta_{\rm sys}\omega_\mathrm{b}$ & $\Delta_{\rm sys}\omega_\mathrm{cdm}$ & $\Delta_{\rm sys}n_s$ & $\Delta_{\rm sys}b$ & $\Delta_{\rm sys}f_\mathrm{NL}$ \\\hline\hline
        $0.7$ & $0.9$  &  2.8$\times 10^{-6}$
 &  -7.8$\times 10^{-9}$
 &  9.2$\times 10^{-7}$
 &  2.3$\times 10^{-6}$
 &  -3.0$\times 10^{-5}$
 &  1.1\\ &
 &  0.0007 $\sigma$
 &  -5.2$\times 10^{-5}$ $\sigma$
 &  0.00093 $\sigma$
 &  0.00061 $\sigma$
 &  -0.0014 $\sigma$
 &  0.015 $\sigma$\\
        \hline
        $0.9$ & $1.1$  &  9.1$\times 10^{-8}$
 &  -2.7$\times 10^{-10}$
 &  3.1$\times 10^{-8}$
 &  7.3$\times 10^{-8}$
 &  -1.1$\times 10^{-6}$
 &  0.019\\  &
 &  2$\times 10^{-5}$ $\sigma$
 &  -1.8$\times 10^{-6}$ $\sigma$
 &  3$\times 10^{-5}$ $\sigma$
 &  2.0$\times 10^{-5}$ $\sigma$
 &  -4.5$\times 10^{-5}$ $\sigma$
 &  0.00049 $\sigma$
 \\\hline
        $1.1$ & $1.3$  &  4.0$\times 10^{-6}$
 &  -1.2$\times 10^{-8}$
 &  1.4$\times 10^{-6}$
 &  3.2-06
 &  -5.1-05
 &  0.51\\ &
 &  0.00098 $\sigma$
 &  -8.1$\times 10^{-5}$ $\sigma$
 &  0.0014 $\sigma$
 &  0.00087 $\sigma$
 &  -0.0020 $\sigma$
 &  0.020 $\sigma$\\
        \hline
        $1.3$ & $1.5$  &  1.4$\times 10^{-5}$
 &  -4.2$\times 10^{-8}$
 &  4.7$\times 10^{-6}$
 &  1.1-05
 &  -0.00019
 &  1.3\\ &
 &  0.0034 $\sigma$
 &  -0.00028 $\sigma$
 &  0.0048 $\sigma$
 &  0.0030 $\sigma$
 &  -0.0068 $\sigma$
 &  0.069 $\sigma$
\\\hline
        $1.5$ & $1.8$ &  6.0$\times 10^{-5}$
 &  -2.0$\times 10^{-7}$
 &  2.1-05
 &  4.7-05
 &  -0.00088
 &  3.3\\ &
 &  0.015 $\sigma$
 &  -0.0013 $\sigma$
 &  0.022 $\sigma$
 &  0.013 $\sigma$
 &  -0.030 $\sigma$
 &  0.30 $\sigma$\\\hline
    \end{tabular}
    \caption{Expected shift of the best-fitting $\Lambda$CDM+$f_\mathrm{NL}$ parameters due to neglecting the effect of the Planck dipole if we measure the parameters in four redshift bins with boundaries $z_\mathrm{min} < z < z_\mathrm{max}$. We use the Gaussianised power spectrum $\Psi(k)$ to account for having only a small amount of ultra-large-scale modes.}
    \label{tab:FS_shifts_in_z_bins}
\end{table}

\subsection{DESI-like Case}
\label{DESI}

\begin{figure}
    \centering
    \includegraphics[width=\textwidth]{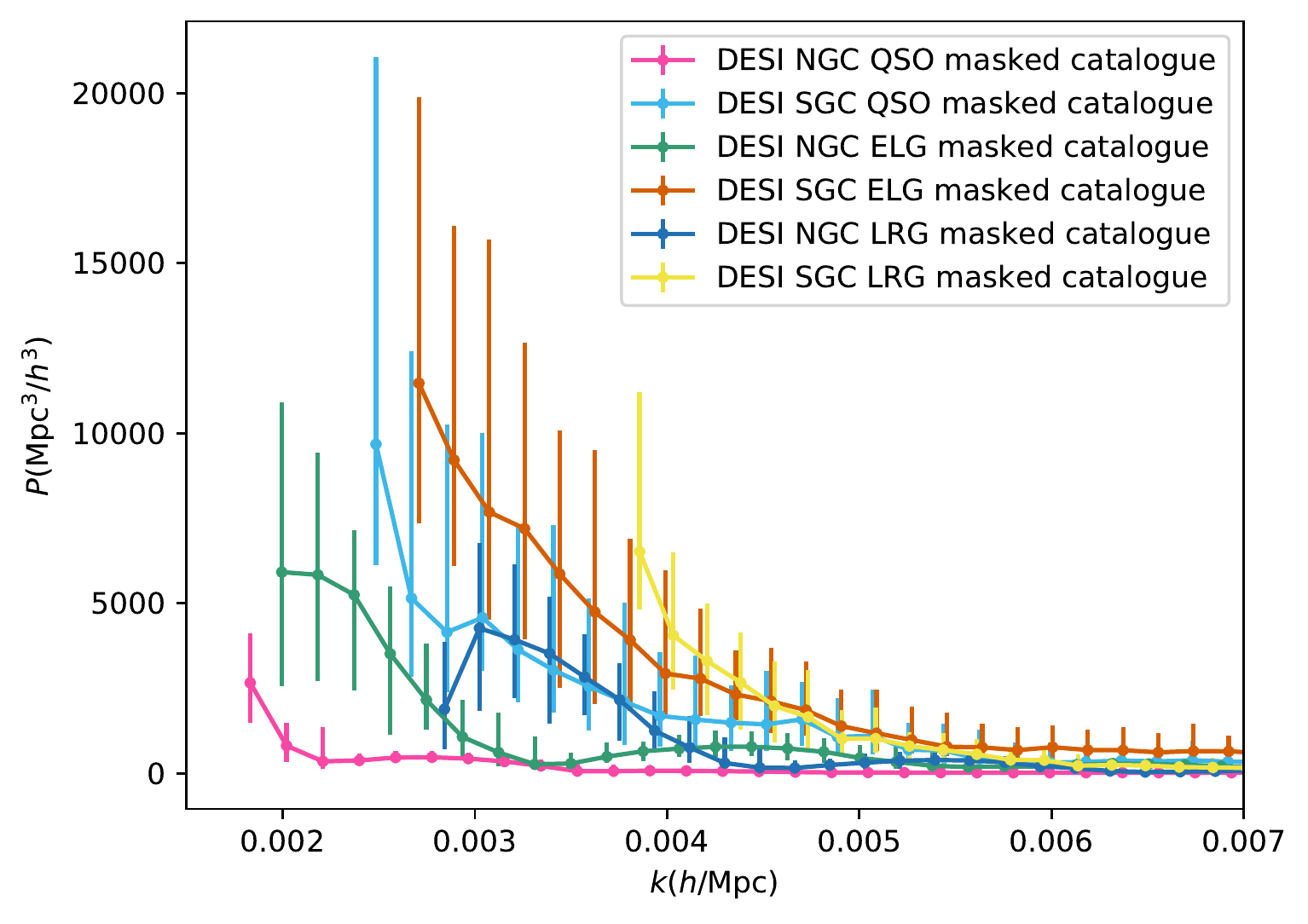}
    \caption{The rocket power spectrum for DESI-like probes.}
    \label{fig:P_DESI}
\end{figure}

\begin{figure}
    \centering
    \includegraphics[width=\textwidth]{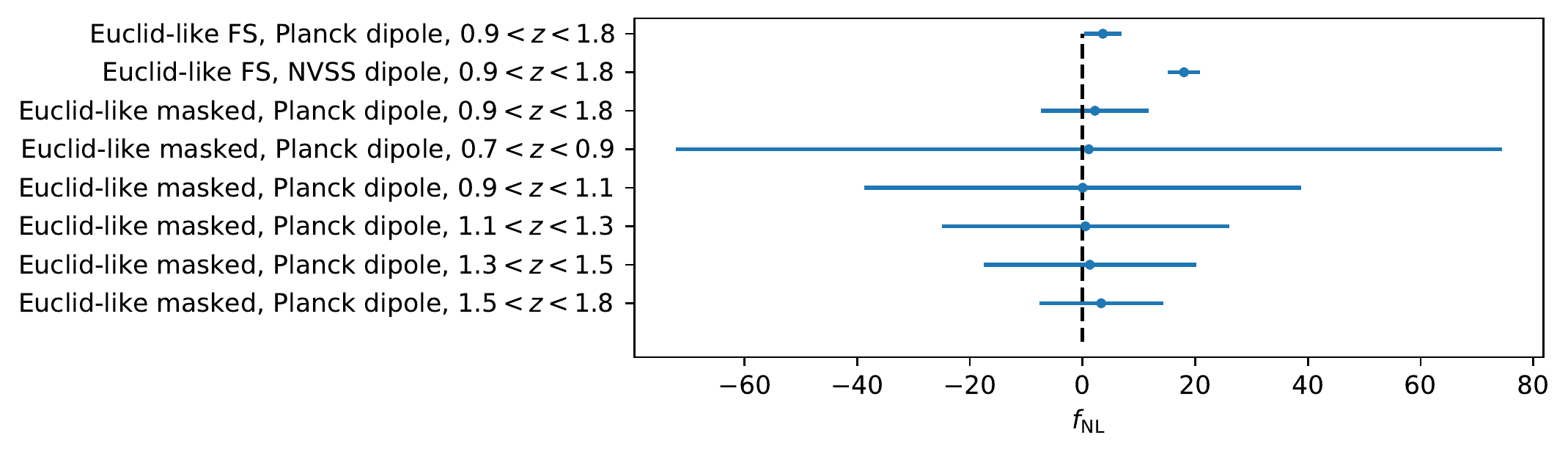}
    \includegraphics[width=\textwidth]{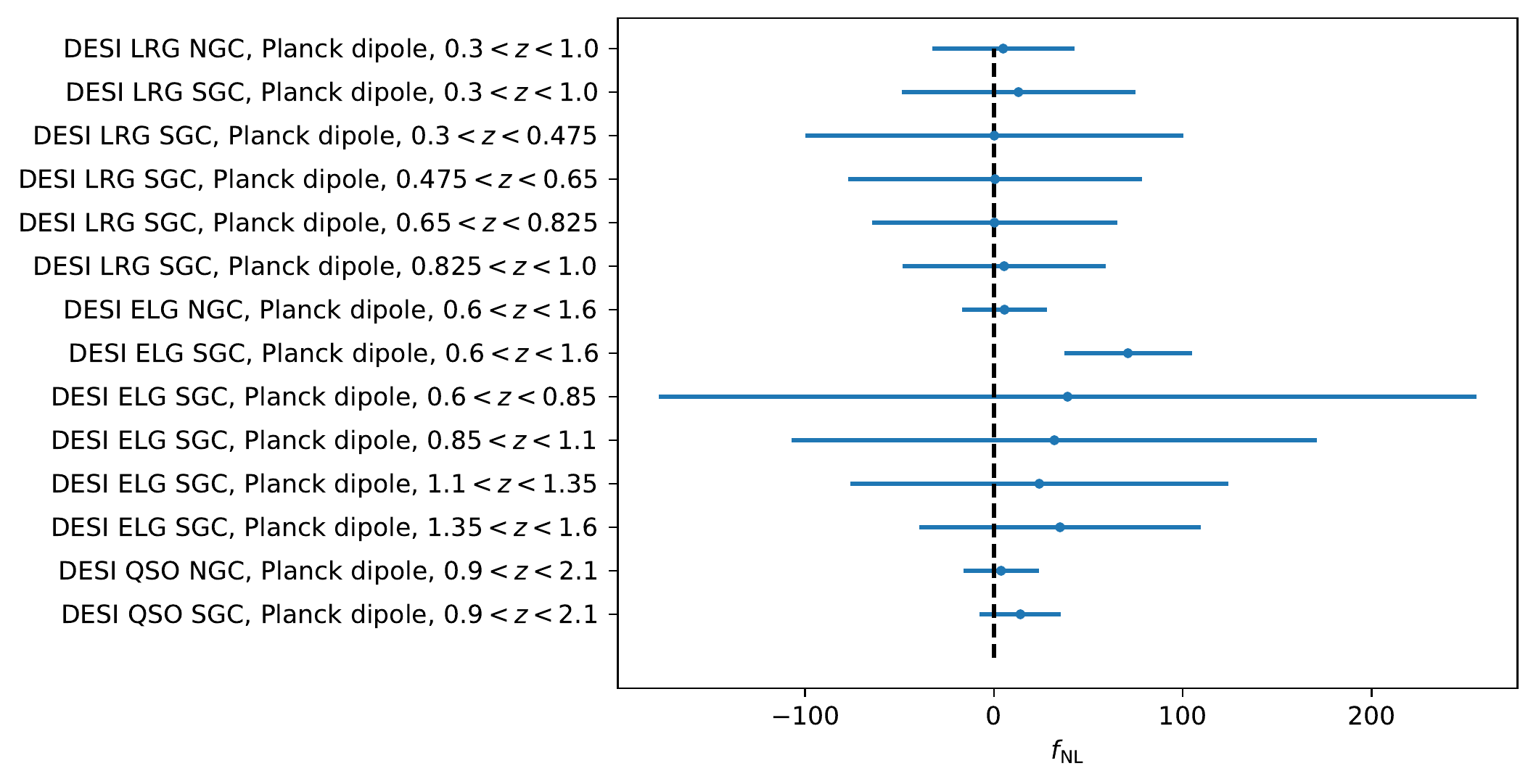}
    \caption{Graphical representation of the shifts of the best-fitting value of $f_\mathrm{NL}$ presented in Tables \ref{tab:FS_shifts_main} - \ref{tab:DESI_tomo}. As in the tables, the error bars correspond to the Cram\'er-Rao bound $\sigma_{f_\mathrm{NL}}=\sqrt{(F^{-1})_{f_\mathrm{NL}f_\mathrm{NL}}}$. Note that all forecasts have been made assuming a fixed bias parameter and growth factor for the entire redshift range covered. Because of that, the forecasts for wide redshift ranges can be significantly different from the mean of the forecasts obtained by binning in smaller sub-ranges. Furthermore, the DESI SGC footprint is aligned with the dipole anti-direction. Therefore, radial modes are crucial. These are missed in the tomographic case for modes that are longer than the width of the tomographic redshift shell, thus, diminishing the Kaiser rocket signal.}
    \label{fig:shift_summary}
\end{figure}

Having established that the effect of the Kaiser rocket depends both on the shape of the radial selection function as well as on the directions covered by the survey mask, we cannot assume that the results obtained so far in this section are representative for all Stage IV spectroscopic galaxy surveys. We therefore continue by forecasting the bias on cosmological parameters measured by a DESI-like survey. For simplicity, we use a circular mask covering an area of 10,000 square degrees centred around $(\mathrm{RA, DEC})=(180,32.5)^\circ$ which approximates the Northern Galactic part of the mask presented in \cite{Dey:2018pfm} well enough for our purposes. In the South, we model the mask as a circle that the cut above $\mathrm{DEC} = 30^\circ$ and below $\mathrm{DEC} = -20^\circ$. The radius is chosen such that the final mask covers an area of 4,000 square degrees. We adopt the bias values $b_\mathrm{QSO} = 1.2/D(z)$, $b_\mathrm{ELG} = 0.84/D(z)$ and $b_\mathrm{LRG} = 1.7/D(z)$ for the quasar, emission line galaxy and luminous red galaxy samples, respectively, from \cite{Aghamousa:2016zmz}. We read the radial selection functions from plots in \cite{Raichoor:2020jcl,Yeche:2020tjm,Zhou:2020mgr} and reproduce them in Figure \ref{fig:dNdz_histograms}. As the amplitude of the Kaiser rocket effect is determined by the log-derivative of the radial selection function, we also show this quantity in the bottom panel of the same figure. Because of the strong dependence of the effect on this quantity and the preliminary nature of the determination of $\bar{N}(z)$ and its derivative, what follows should be considered more like a proof of principle than an accurate forecast. We can notice that the selection function of the DESI quasar sample is very smooth over a wide range of redshifts and its log-derivative is therefore smaller than the one of the Euclid-like galaxy sample. We therefore expect the Kaiser rocket effect to not affect cosmological measurements using the DESI quasar sample significantly. On the other hand, the selection functions of emission line galaxies (ELG) and luminous red galaxies (LRG) has many features which means that accounting for the Kaiser rocket effect may be relevant when doing inference with these samples. When analyzing ELG in redshift bins  we might not 
see the aforementioned telltale sign of $f_\mathrm{NL}$ varying with redshift due to multiple regularly spaced saddle points in its selection function. On the other hand, the more complicated selection function shape results in power spectrum k-dependence that is  easier to distinguish from an $f_\mathrm{NL}$ signal (cf. Figure \ref{fig:P_DESI}). Combined with the fact that DESI LRG  and Quasi Stellar Objects (QSO) are biased more strongly than Euclid-like objects, the shift in the best-fitting value of $f_\mathrm{NL}$ is smaller in these two probes. However, for the ELG sample, we obtain a result in the Galactic North that is comparable to the shifts we expect for a Euclid-like survey, and in the South we even expect a $2\;\sigma$ bias in $f_\mathrm{NL}$. We list the expected shifts in cosmological parameters in Table \ref{tab:FS_shifts_masked_DESI} and plot them in Figure \eqref{fig:shift_summary}. 

As the DESI ELG and LRG radial selection functions are less smooth than the DESI QSO and Euclid-like selection functions, we repeat the tomographic analysis where we split the whole sample into four redshift bins. As the effect is stronger in the South than in the North, we only present results for the South in Table \ref{tab:DESI_tomo} and Figure \ref{fig:shift_summary} and in the right panel of Figure \ref{fig:param_shifts_in_z_bins}. As was the case for the Euclid-like $N(z)$, we find also for these more complicated selection functions that the parameter bias vanishes as $\d\ln N/\d z$ vanishes. Furthermore, the bias in $f_\mathrm{NL}$ is less in the tomographic analysis than when the whole data set is analysed at once. 
Note that all forecasts have  been  made  assuming  a  fixed  bias  parameter  and  growth  factor  for  the  entire  redshift  range  covered. Because  of  that,  the  forecasts  for  wide  redshift  ranges  can  be  significantly  different  from  the  mean  of  the forecasts obtained by binning in smaller sub-ranges. Furthermore, the DESI SGC footprint is aligned with the dipole anti-direction. Therefore, radial modes are crucial. These are missed in the tomographic case for modes that are longer than the width of the tomographic redshift shell, thus, diminishing the Kaiser rocket signal.

\begin{table}[ht]
    \centering
    \begin{tabular}{|l|c|c|c|c|c|c|c|}
        \hline  tracer & hemi- & $\Delta_{\rm sys}$ &$\Delta_{\rm sys}$ &$\Delta_{\rm sys}$ &$\Delta_{\rm sys}$ &$\Delta_{\rm sys}$ &$\Delta_{\rm sys}$ \\
        & sphere & $h$ & $\omega_\mathrm{b}$ & $\omega_\mathrm{cdm}$ & $n_s$ & $b$ & $f_\mathrm{NL}$ \\\hline\hline
LRG & NGC   &  2.2e-5
 &  -2.0e-8
 &  6.3e-6
 &  1.8e-5
 &  -0.00041
 &  4.9\\ &
    &  0.0050 $\sigma$
 &  -0.00013 $\sigma$
 &  0.0062 $\sigma$
 &  0.0046 $\sigma$
 &  -0.011 $\sigma$
 &  0.13 $\sigma$\\
 $b = 2.4$ & SGC  &  9.6$\times 10^{-5}$
 &  -2.0$\times 10^{-7}$
 &  3.1$\times 10^{-5}$
 &  8.034$\times 10^{-5}$
 &  -0.0017
 &  13\\ &
 &  0.020 $\sigma$
 &  -0.0014 $\sigma$
 &  0.029 $\sigma$
 &  0.020 $\sigma$
 &  -0.039 $\sigma$
 &  0.21 $\sigma$\\
 \hline
 ELG & NGC  &  2.4e-5
 &  -6.7e-7
 &  1.7e-5
 &  5.4e-5
 &  -0.00049
 &  5.6\\
 &   &  0.0097 $\sigma$
 &  -0.00503 $\sigma$
 &  0.018 $\sigma$
 &  0.015 $\sigma$
 &  -0.035 $\sigma$
 &  0.25 $\sigma$\\
$b = 1.5$  & SGC  &  0.00065
 &  -2.6$\times 10^{-6}$
 &  0.00025
 &  0.00058
 &  -0.0079
 &  71\\
 &   &  0.14 $\sigma$
 &  -0.017 $\sigma$
 &  0.24 $\sigma$
 &  0.15 $\sigma$
 &  -0.30 $\sigma$
 &  2.1 $\sigma$\\\hline
 QSO  & NGC  &  1.5e-5
 &  6.3e-9
 &  3.5e-6
 &  9.1e-6
 &  -0.00063
 &  3.8 \\ &  &  0.0030 $\sigma$
 &  4.2e-5 $\sigma$
 &  0.0031 $\sigma$
 &  0.0022 $\sigma$
 &  -0.013 $\sigma$
 &  0.19 $\sigma$\\
$b = 2.4$ & SGC  &  0.00026
 &  -5.8$\times 10^{-7}$
 &  7.4$\times 10^{-5}$
 &  0.00018
 &  0.013
 &  14\\
 &  &  0.049 $\sigma$
 &  -0.0039 $\sigma$
 &  0.064 $\sigma$
 &  0.044 $\sigma$
 &  0.25 $\sigma$
 &  0.65 $\sigma$\\\hline
    \end{tabular}
    \caption{Expected shift of the best-fitting $\Lambda$CDM+$f_\mathrm{NL}$ parameters similar to Table \ref{tab:FS_shifts_main} but for a masked DESI-like survey assuming that the bulk motion is described by the CMB dipole measured by Planck.}
    \label{tab:FS_shifts_masked_DESI}
\end{table}

\begin{table}[ht]
    \centering
    \begin{tabular}{|l|l|l|c|c|c|c|c|c|c|}
        \hline  tracer & $z_\mathrm{min}$ & $z_\mathrm{max}$ & bias & $\Delta_{\rm sys}$ &$\Delta_{\rm sys}$ &$\Delta_{\rm sys}$ &$\Delta_{\rm sys}$ &$\Delta_{\rm sys}$ &$\Delta_{\rm sys}$ \\ & &
        &  & $h$ & $\omega_\mathrm{b}$ & $\omega_\mathrm{cdm}$ & $n_s$ & $b$ & $f_\mathrm{NL}$ \\\hline\hline
        ELG & 0.6 & 0.85 & 1.32  &  -0.00036
 &  -2.8e-7
 &  -8.0e-5
 &  -0.00016
 &  0.0030
 &  39\\ & & & &  0.073 $\sigma$
 &  -0.0019 $\sigma$
 &  -0.070 $\sigma$
 &  -0.041 $\sigma$
 &  0.10 $\sigma$
 &  0.18 $\sigma$\\
 & 0.85 & 1.1 & 1.44  &  -0.00034
 &  -3.2e-7
 &  -7.3e-5
 &  -0.00015
 &  0.0030
 &  32\\ & & & 
  &  -0.069 $\sigma$
 &  -0.0022 $\sigma$
 &  -0.063 $\sigma$
 &  -0.037 $\sigma$
 &  0.094 $\sigma$
 &  0.23 $\sigma$\\
 & 1.1 & 1.35 & 1.57 &  -0.00036
 &  -3.32e-7
 &  -7.8e-5
 &  -0.00016
 &  0.0034
 &  24\\ & & &
  &  -0.074 $\sigma$
 &  -0.0022 $\sigma$
 &  -0.068 $\sigma$
 &  -0.040 $\sigma$
 &  0.10 $\sigma$
 &  0.24 $\sigma$\\
 & 1.35 & 1.6 & 1.69 &   -0.00072
 &  -6.7e-7
 &  -0.00016
 &  -0.00032
 &  0.0074
 &  35\\ & & & 
  &  -0.15 $\sigma$
 &  -0.0045 $\sigma$
 &  -0.14 $\sigma$
 &  -0.080 $\sigma$
 &  0.20 $\sigma$
 &  0.47 $\sigma$
\\\hline
LRG & 0.3 & 0.475 & 2.22 &  1.9e-7
 &  -7.1e-10
 &  6.7e-8
 &  1.9e-7
 &  -4.9e-6
 &  0.17\\ & & & &  3.9e-5 $\sigma$
 &  -4.7e-6 $\sigma$
 &  6.1e-5 $\sigma$
 &  4.8e-5 $\sigma$
 &  -1.2e-4 $\sigma$
 &  0.0017 $\sigma$\\
 & 0.475 & 0.65 & 2.38  &  7.5e-7
 &  -1.7e-9
 &  2.4e-7
 &  6.9e-7
 &  -2.0e-5
 &  0.56\\ & & &  &  1.5e-4 $\sigma$
 &  -1.2e-5 $\sigma$
 &  2.1e-4 $\sigma$
 &  1.7e-4 $\sigma$
 &  -4.7e-4 $\sigma$
 &  0.0072 $\sigma$\\
 & 0.65 & 0.825 & 2.54  &  5.7e-7
 &  -1.0e-9
 &  1.7e-7
 &  4.7e-7
 &  -1.5e-5
 &  0.26\\ & & & &  1.2e-4 $\sigma$
 &  -6.7e-6 $\sigma$
 &  1.5e-4 $\sigma$
 &  1.2e-4 $\sigma$
 &  -3.1e-4 $\sigma$
 &  0.0040 $\sigma$\\
 & 0.825 & 1.0 & 2.71  &  -3.1e-5
 &  4.6e-8
 &  -1.0e-5
 &  -3.2e-5
 &  0.0026
 &  5.4\\ & & &  &  -0.0065 $\sigma$
 &  3.1e-4 $\sigma$
 &  -0.0093 $\sigma$
 &  -0.0080 $\sigma$
 &  0.051 $\sigma$
 &  0.10 $\sigma$\\\hline
    \end{tabular}
    \caption{Expected shift of the best-fitting $\Lambda$CDM+$f_\mathrm{NL}$ parameters similar to Table \ref{tab:FS_shifts_masked_DESI} but for a masked DESI-like ELG and LRG survey covering the Southern Galactic Cap in tomographic redshift bins.}
    \label{tab:DESI_tomo}
\end{table}

\section{Testing the assumption of uncorrelated  cosmology and rocket signal on Gaussian Random Fields}
\label{sec:GRF}
One of the basic assumptions in  our modelling of the Kaiser rocket signal as a contaminant to the power spectrum (cf. Section \ref{sec:meth}) is that the signal from the cosmological background and the one from the Kaiser rocket effect are uncorrelated.
This is not the case in detail, because the LG velocity  causing the effect is sourced by local structure. However we expect this correlation to be small, especially  at large scale,   where  power in different k-modes is uncorrelated.
This is what we demonstrate here.
\subsection{Generating cosmological Gaussian random fields}
As we are interested in very large, linear $k$-modes, we can  resort to realisations of Gaussian random fields. To generate Gaussian realisations of $\delta_\mathrm{cosmo}$, we proceed as standard by drawing real and imaginary parts of $\delta({\bf k})$ from zero-mean Gaussians with variance $L^3P(k)/2$, where $L$ is the length of the cubic box and $P(k)$ is the power spectrum of the fiducial cosmology.
We ensure that integral constraint, and  reality of the $\delta({\bf r})$  are satisfied, in particular we impose $\delta(-{\bf k})=\delta^*({\bf k})$, and use  discrete Fourier transform routine fftw \cite{FFTW05}\footnote{Fastest Fourier Transform in the West: http://fftw.org} to obtain the configuration space realisation $\delta_\mathrm{cosmo}(\mathbf{r})$.
Then within each grid cell, we place $\bar n V_\mathrm{cell}(1 + \delta_\mathrm{cosmo}(\mathbf{r}))$ objects at random positions within the cell volume $V_\mathrm{cell}$.
Finally, after suitably placing the observer, we transform distances into  redshifts (and angular positions) according to the fiducial cosmology and  downsample the  objects such that the final catalogue has the  required redshift distribution $N(z)$ and angular mask corresponding to  the survey footprint.

\subsection{Testing the Independence between Cosmological and Kaiser Rocket Induced Signal}
\label{sec:crosstest}

 Having generated many Gaussian random field realisations, we now test the independence of the two signals. Taking advantage of the linearised continuity equation
\begin{equation}
    \frac{\partial\delta_\mathrm{cosmo}(\mathbf{r})}{\partial t} + \nabla.\mathbf{v}(\mathbf{r}) = 0,
    \label{eq:conti}
\end{equation}
we  obtain the velocity field $\mathbf{v}(\mathbf{r})$ corresponding to the density field realisation $\delta_\mathrm{cosmo}(\mathbf{r})$ going to Fourier space, where eq. \eqref{eq:conti} reduces to 
\begin{align}
    \mathbf{v}(\mathbf{k}) &= afH\frac{i\mathbf{k}}{k^2}\frac{\delta_\mathrm{cosmo}(\mathbf{k})}{b}\text{, if }\mathbf{k}\neq 0\text{, and}\\
    \mathbf{v}({\bf 0}) &= 0,\nonumber
\end{align}
where $a$ is the scale factor. After Fourier transforming $\mathbf{v}$ to configuration space, we search in each realisation for the position of the most \textit{Planck}-like observer. We do so by minimising $\vert \mathbf{v} - \mathbf{v}_\mathrm{Planck}\vert + w_1\vert \mathbf{x}\vert + w_2\left(\vert RA - RA_\mathrm{Planck}\vert + \vert DEC - DEC_\mathrm{Planck}\vert\right)$, where the second term penalises observers that are too far from the simulation box and therefore would look outside of it and the second term makes sure that the dipole direction of the observer $(RA, DEC)$ is similar to the Planck one $(RA_\mathrm{Planck}, DEC_\mathrm{Planck})$, which will be important when considering the survey mask later. We weight the importance of these terms with weights that we choose to be  $w_1 = 10^{-9}h\;\mathrm{Mpc}^{-1}$ and $w_2 = 10^{-2}\;\mathrm{deg}^{-1}$. We thus place the observer, and then apply the selection as described above and compute the power spectrum of this selected catalogue. After that, we shift the objects in the selected catalogue according to the dipole motion that the observer experiences and measure the power spectrum of these shifted objects. To estimate the cross-correlation, we repeat the same procedure but positioning the observer at a random position and shifting the objects according to the Planck dipole.

\begin{figure}
    \centering
    \includegraphics[width=\textwidth]{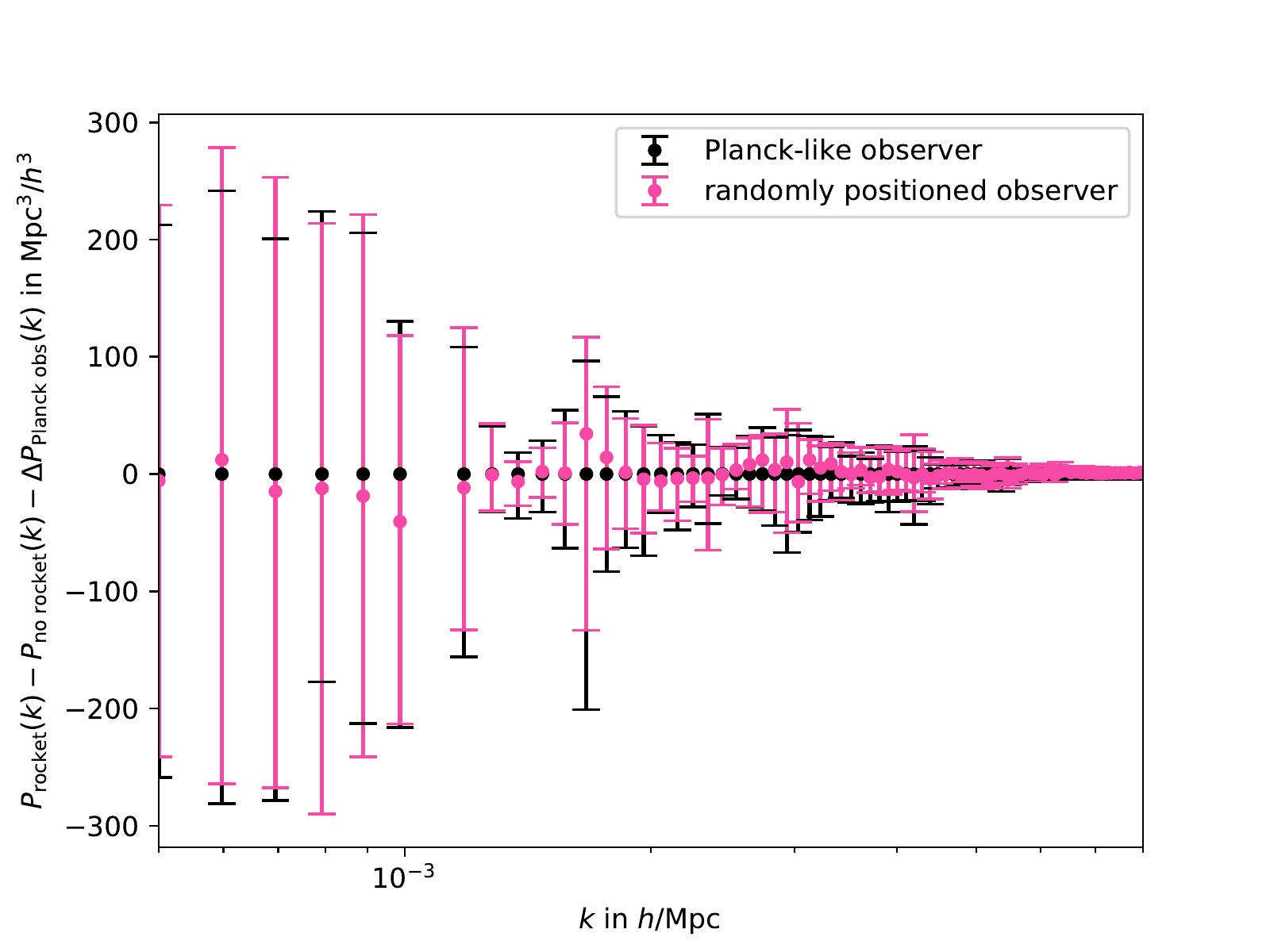}
    \caption{Residual (compared to the Planck-like observer case) of the Power spectra of the difference in density between shifted and unshifted objects of observers placed at random positions (magenta) and those placed at positions where the absolute value of the velocity field is the most similar to the dipole amplitude measured by Planck (black). The error bars have to be understood as for a single realisation since that is what we can observe in reality.}
    \label{fig:KaiserRocket_power_FS_from_GRF}
\end{figure}

We plot the median power spectrum of the difference in density between shifted and unshifted objects in $N_\mathrm{realisations} = 100$ realisations in Figure \ref{fig:KaiserRocket_power_FS_from_GRF}. Using the 16- and 84-percentiles to estimate the confidence intervals $I_\mathrm{conf}$ of the difference in power in a single realisation, and multiplying the $I_\mathrm{conf}$ with $1.25$ to obtain the confidence interval on the median of a single realisation. We  observe consistency between the power spectrum excess induced by the Kaiser rocket effect where the peculiar velocity agrees with the density distribution of the simulation and  that where the dipole velocity is set by hand at a random position.
This indicates that the correlation between the Kaiser rocked induced signal and the cosmological signal is negligible, hence it is safe to assume that $P(k)=P_\mathrm{cosmo}(k)+ P_\mathrm{rocket}(k)$.

\section{Accounting for the Kaiser Rocket Effect}
 \label{sec:howtoaccountfor}
 We can recognise at least three ways to account for a Kaiser rocket effect in a realistic survey. As discussed in Section \ref{sec:model} the most transparent and elegant way would be using the spherical-Bessel expansion, modelling the rocket effect as a dipolar signal uniquely determined by the derivative of the number density,  and the orientation and amplitude of the observer's velocity.
 This approach is practical at very large scales; at small scales it becomes numerically very intensive motivating hybrid approaches such as Ref.~\cite{Mike:2018zvb}.
 Another approach involves correcting for this signal in the catalogues themselves or in the random catalogues, which are a key ingredient of any FKP-type  weighting, and are used already to remove spurious signals. This approach is straight-forward if the dipole amplitude and direction are precisely known. However, it becomes  expensive if marginalised, especially over direction. A  less numerically intensive technique, which allows for fast marginalization over the amplitude of the signal is based on constructing a suitable template.

 Our model of the observed density field 
 \begin{equation}
     \delta_\mathrm{obs}(\mathbf{x}) = \delta_\mathrm{cosmo}(\mathbf{x}) + \delta_\mathrm{rocket}(\mathbf{x}) = \delta_\mathrm{cosmo}(\mathbf{x}) + v \cos\theta\left[1 + (1 + z)\frac{\d \ln \bar N(z)}{\d z}\right]
     \label{eq:rocket_as_contaminant}
 \end{equation}
 coincides with the underlying model of contaminants in mode-deprojection techniques described in \cite{Slosar:2004fr,Ho:2008bz,Pullen:2012rd,Elsner:2015aga,Kalus:2016cno,Elsner:2016bvs,Kalus:2018qsy}. The idea behind mode deprojection \cite{Rybicki:1992jz} is that if a contaminant can be described by a template $f$, or a set of orthogonal templates $\lbrace f_i \rbrace$, with unknown amplitude, one can marginalise out the contaminated modes by replacing the covariance between Fourier modes $\mathbf{k}_\alpha$ and $\mathbf{k}_\beta$ by
 \begin{equation}
     C_{\alpha\beta} \rightarrow C_{\alpha\beta} + \lim_{\sigma\rightarrow\infty}\sigma f(\mathbf{k}_\alpha)f^\ast(\mathbf{k}_\beta)
 \end{equation}
 in a covariance-based estimator of the power spectrum, such as the Quadratic Maximum Likelihood Estimator \cite[QML]{Tegmark:1996qt}. As we have argued earlier, the amplitude $v$ of our motion dipole is not as well established at LSS redshifts as at the CMB. Therefore, we can marginalise over $v$ if our template $f(\mathbf{k})$ is chosen as the Fourier transform of 
\begin{equation}
     f(\mathbf{x}) = \cos\theta\left[1 + (1 + z)\frac{\d \ln \bar N(z)}{\d z}\right].
 \end{equation}
 Alternatively, assuming that $\delta_\mathrm{cosmo}$ is Gaussian, one can find the best-fitting amplitude of the motion dipole amplitude $v_\mathrm{BF}$ and, hence, an estimate
 \begin{equation}
     \delta_\mathrm{cosmo}^\mathrm{(BF)} = \delta_\mathrm{obs} - v_\mathrm{BF}f
     \label{eq:linear_contamination}
 \end{equation}
 of the cosmic density field. It has to be emphasised that 
 $\langle \vert \delta_\mathrm{cosmo}^\mathrm{(BF)} \vert^2\rangle$
 yields a biased estimate of the power spectrum because $\delta_\mathrm{obs}$ and $v_\mathrm{BF}$ are correlated \cite{Elsner:2015aga}. In \cite{Kalus:2016cno}, a debiasing step is provided that is implemented in the MOde Subtraction code to Eliminate Systematic contamination in galaxy clustering power spectrum measurements \cite[MOSES]{Kalus:2018qsy}.\footnote{https://github.com/KalusB/Moses}

 Having tested our basic assumption of independence between the cosmological and Kaiser rocket effect induced power spectrum contributions, we continue using the same Gaussian random fields to test some of the strategies for removing the spurious clustering signal induced by the Kaiser rocket effect presented in Section \ref{sec:howtoaccountfor} in 3D power spectrum analyses. To be more realistic, we only use the realisations where the observer has been placed in positions such that both the amplitude and the direction of the velocity field are similar (in terms of the definition given in the previous subsection) to the CMB dipole measured by Planck \cite{Akrami:2018vks}.

\begin{figure}
     \centering
     \includegraphics[width=\textwidth]{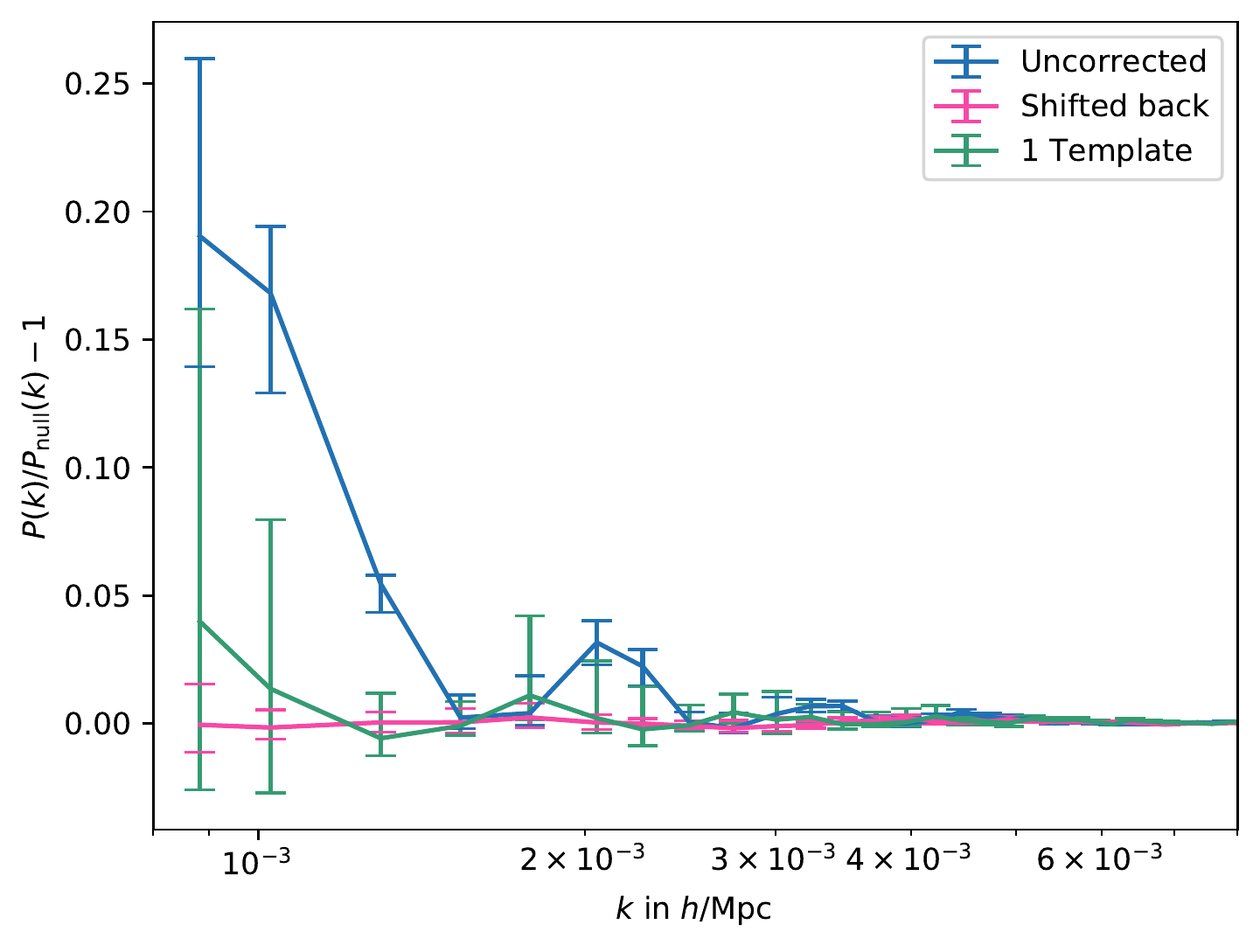}
    \caption{The median and $1\sigma$-confidence intervals of the residual power spectra of 100 Gaussian random field realisations. The residuals are with respect to the "Null" spectrum which is the power spectrum measured from each realisation before applying the Kaiser dipole shift to the catalogue. The blue "Uncorrected" data points and lines refer to the power spectra obtained after shifting the objects in the catalogues without applying any sort of correction to them. The magenta "Shifted back" values have been obtained from catalogues where objects are shifted back assuming that the observer's local dipole motion corresponds exactly to the CMB dipole measured by the Planck satellite \cite{Akrami:2018vks}. Finally, for the green lines, we use Moses \cite{Kalus:2018qsy} to subtract contaminated modes using a Kaiser rocket template and marginalising over its unknown amplitude.}
     \label{fig:cleaning}
 \end{figure}

We compare in Figure \ref{fig:cleaning} the power spectra that we obtain with different mitigation strategies to a null power spectrum obtained from each catalogue  before the apparent shifts due to the observer bulk motion are applied. 

 Our starting point is the power spectrum of the catalogues after shifting each object according to its angular separation with respect to the direction of the bulk motion. We can see in Figure \ref{fig:cleaning} that the Kaiser rocket effect causes an excess of about 20 per cent in the power spectrum in the first $k$-bin. As expected, the power spectrum measured by observers in motion is inconsistent with the "null" power spectrum up to $k\sim 0.007 h/\mathrm{Mpc}$. Above this scale, the observer's bulk motion does not affect their clustering measurements.

 The most na\"ive way to account for the Kaiser rocket effect is to assume an amplitude and a direction of the observer's motion dipole and shift back every object in the galaxy catalogue. As we are performing the same shift to several catalogues, it is more efficient, but mathematically equivalent, to forward shift the random catalogue used in computing the power spectrum. In this way, the null power spectrum can be reconstructed, as expected, in this simple case where the dipole is exactly known. 
 
 However, up to 33 per cent of the CMB dipole can be due to effects other than our local motion \cite{Saha:2021bay}, and local measurements of the radio dipole \cite{Blake:2002gx,Singal:2011dy,Gibelyou:2012ri,Rubart:2013tx,Kothari:2013gya,Tiwari:2013ima,Tiwari:2015tba} do not agree with the amplitude of the CMB dipole. To that end, we run MOSES with a Kaiser rocket template, thus marginalising over all possible values of the amplitude of the observer's peculiar velocity, to obtain the green power spectrum. The template-based mitigation result is also consistent with the null power spectrum, but, expectedly, with a much larger uncertainty at the scales most affected by the Kaiser rocket. However, this faithfully reflects our uncertainty of the peculiar motion measured from local data.

 \label{sec:testing_mitigation}
Our simplifying assumption of a mask  separable in radial and angular components may be insufficiently accurate in a realistic application. We argue here that as long as the full dependence (both angular and radial) of the mask is known, it can be accounted for in the random catalogues. Mitigation strategies based on the random catalogues can therefore  account for  this complication in a straightforward way. On the other hand,  random-catalogue-based mitigation strategies make marginalization over uncertain quantities (dipole velocity and direction) more (possibly even prohibitively) computationaly intensive.  The implications for  other mitigation approaches goes beyond  the scope of this paper, but one possible approach would be to  decompose the  window as a leading order separable contribution and  sub-dominant non-separable corrections.  Thus the rocket effect will have a leading contribution and a correction. The marginalization can be carried out on the leading order contribution and the correction terms could be evaluated  for fiducial values of the  uncertain quantities.
It is important to note that different surveys with different $d\ln N/dz$ should see the same dipole hence the same velocity and direction (as long as their depths are comparable or convergence to the cosmological dipole has been reached) but the rocket signal should be modulated by the specific $d\ln N/dz$. This offers a powerful  consistency check.

\section{Conclusions} 
\label{sec:conclusions}

The peculiar motion of the observer, if not fully and accurately  accounted for, induces a well known clustering signal in the galaxy distribution which is related to the Kaiser Rocket effect. This dipole due to our motion with respect to a rest frame where the galaxy distribution is statistically isotropic is expected to converge to the CMB one if the galaxy survey is deep enough and the cosmological principle holds. 

The rocket-induced  spurious clustering is superimposed to the cosmological one and, we show, largely uncorrelated. When computing the workhorse summary statistics, the  3D (Cartesian coordinates) power spectrum, the measured power spectrum signal is given by the superposition (addition) of the cosmological and the rocket induced one. The rocket power spectrum is scale dependent, relevant at ultra-large scales and can bias the estimation of cosmological parameters. Of particular concern is a possible systematic bias on the local non-Gaussianity parameter. $f_\mathrm{NL}$. Through analytical calculations, mock simulations and Fisher forecasts we have found that for  realistic surveys, and realistic dipole directions and amplitude the bias induced in cosmological parameters is small except for the primordial non-Gaussianity parameter of the local type.  In this case for surveys selection functions where the galaxy number density varies steeply with redshift (large $\d N/\d z$) the induced bias can reach the $2-\sigma$ level.  However this effect , and the subsequent systematic bias in the recovered parameters, is zero where $\d N/\d z$ is zero. This is an internal consistency check that helps detecting a possible rocket effect and disentangle this spurious signal from a new physics signal. For such survey configuration, the detection and measurement of a dipole due to our motion with respect to a rest frame where the galaxy distribution is statistically isotropic could be used to test the cosmological principle (see also \cite{Maartens:2017qoa}).  A robust measurement of a galaxy dipole inconsistent with the CMB dipole from forthcoming (wide angle and covering a large redshift range) galaxy surveys would be a smoking gun for new physics, right at the basic pillars on which the standard cosmological model is built. 

Even if the rocket-induced galaxy dipole is  consistent with the CMB one, and  for most survey configurations 
the rocket induced power spectrum is small and its  systematic effect on parameter estimation is small,  we argue that it is a well known systematic effect which should thus be removed. 
We have presented several ways to do so ranging from suitably modulating the random catalogues which are routinely used to compute the power spectrum estimator of choice, to template based approaches.

Some mitigation techniques are most suitable if the amplitude and direction of the dipole are perfectly known, others are better suited if the uncertainty in the dipole amplitude (recall that the NVSS dipole amplitude is a factor of 3 larger than the CMB one yet with consistent direction) should be propagated into cosmological inference.  

Before concluding, some caveats are in order. 
The natural basis to describe the rocket effect is the spherical Fourier-Bessel basis. Most of the complications induced by the rocket effect described in this paper would be easier to address, but it would complicate significantly the  interpretation of the cosmological signal  at scales significantly smaller than the survey size.  
Apart from the issue of non-Gaussian two-point statistics, at the ultra-large scales of interest for $f_\mathrm{NL}$ constraints, wide-angle and relativistic effects are critical. The former is difficult to implement in the standard estimators used commonly in LSS analysis. For instance, the popular Yamamoto estimator \cite{Yamamoto:2005dz} is not suitable when dealing with large angular separations since it still assumes a single line of sight. Furthermore, the gravitational lensing caused by foreground objects may result in a dilution of the observed number of objects in a given patch of the sky. To compute $P(k)$, we have to perform a Fourier transform on a three-dimensional hypersurface that is not well-defined due to light-cone dependence of the density field brought about by gravitational lensing. Ignoring gravitational lensing at ultra-large scales in our Euclid-like mock survey shifts the best-fitting $f_\mathrm{NL}$ by about $1\sigma$ \cite{Bernal:2020pwq} and should, therefore, also be taken into account when correcting for the Kaiser rocket effect. For these reasons, we suggest using configuration space observables such as the two-point correlation function $\xi(x)$,\footnote{For a fully relativistic expression of $\xi(x)$ that takes the Kaiser rocket effect into account, we refer to \cite{Bertacca:2019wyg}.} or to infer cosmological parameters from the map directly when working with ultra-large scales. Working in configuration space enables one to better keep track of light cone  and wide angle effects, which can’t be neglected at the large  scales where the dipole signal is relevant.
But if the three dimensional power spectrum is used, its interpretation should be guided by the findings presented here and eq. \eqref{eq:rocket_as_contaminant} should be used to mitigate the Kaiser rocket effect.
In particular, it is important to note that rocket effect signature in the data could be significantly boosted (possibly even up to a factor of $\sim$ few)  for luminosity-limited samples for which the magnification bias is large. This caveat is particularly relevant for the significance and robustness of a constraint or detection of primordial non-Gaussianity.

\appendix
\section{Choosing an appropriate $k_\mathrm{max}$ for our Fisher analyses.}
\label{app:1}
To pick the minimum scale that should enter the Fisher-based analysis of section \ref{sec:survey_mocks}, we perform a simple test analysis for different $k_\mathrm{max}$ and list the results in Table \ref{tab:Euclid_FS_shifts}.  We obtain very similar estimates for the best-fitting parameter shifts for both $k_\mathrm{max}=0.2\;h/\mathrm{Mpc}$ and $k_\mathrm{max}=0.3\;h/\mathrm{Mpc}$, justifying our choice of  $k_\mathrm{max}=0.2\;h/\mathrm{Mpc}$. 
\begin{table}[ht]
    \centering
    \begin{tabular}{|l|c|c|c|c|c|c|c|}
        \hline dipole & $k_\mathrm{max}$ & $h$ & $\omega_\mathrm{b}$ & $\omega_\mathrm{cdm}$ & $n_s$ & $b$ & $f_\mathrm{NL}$ \\
        ampl. & ($h$/Mpc)&$\Delta_{\rm sys}$&$\Delta_{\rm sys}$&$\Delta_{\rm sys}$&$\Delta_{\rm sys}$&$\Delta_{\rm sys}$&$\Delta_{\rm sys}$\\ \hline\hline
        Planck & 0.03  &  2.1$\times 10^{-6}$
 &  -5.6$\times 10^{-8}$
 &  2.0$\times 10^{-6}$
 &  3.6$\times 10^{-6}$
 &  -0.0011
 &  1.4\\
        &  &  0.00039 $\sigma$
 &  -0.00037 $\sigma$
 &  0.0017 $\sigma$
 &  0.00086 $\sigma$
 &  -0.031 $\sigma$
 &  0.12 $\sigma$\\
        & 0.2 &  1.2$\times 10^{-5}$
 &  -2.4$\times 10^{-7}$
 &  1.1$\times 10^{-5}$
 &  2.8$\times 10^{-5}$
 &  -0.00028
 &  1.1
\\
          & & 0.0048 $\sigma$
 &  -0.0016 $\sigma$
 &  0.012 $\sigma$
 &  0.0092 $\sigma$
 &  -0.019 $\sigma$
 &  0.10 $\sigma$\\
        & 0.3  &  2.0$\times 10^{-6}$
 &  -4.4$\times 10^{-7}$
 &  1.1$\times 10^{-5}$
 &  3.0$\times 10^{-5}$
 &  -0.00021
 &  1.1\\
        &  &  0.0011 $\sigma$
 &  -0.0088 $\sigma$
 &  0.012 $\sigma$
 &  0.012 $\sigma$
 &  -0.020 $\sigma$
 &  0.10 $\sigma$\\\hline
    \end{tabular}
    \caption{Expected shift of the best-fitting $\Lambda$CDM+$f_\mathrm{NL}$ parameters due to the Kaiser rocket effect measured from a Euclid-like full-sky survey in a single redshift bin. We present the shift both as an absolute number as well as in terms of the expected uncertainty of each parameter estimated as $\sigma(\vartheta_\alpha)=\sqrt{(F^{-1})_{\alpha\alpha}}$.}
    \label{tab:Euclid_FS_shifts}
\end{table}

\section{Gaussian versus Gaussianised Power Spectrum}
\label{G_vs_nG}
\begin{figure}
    \centering
    \includegraphics{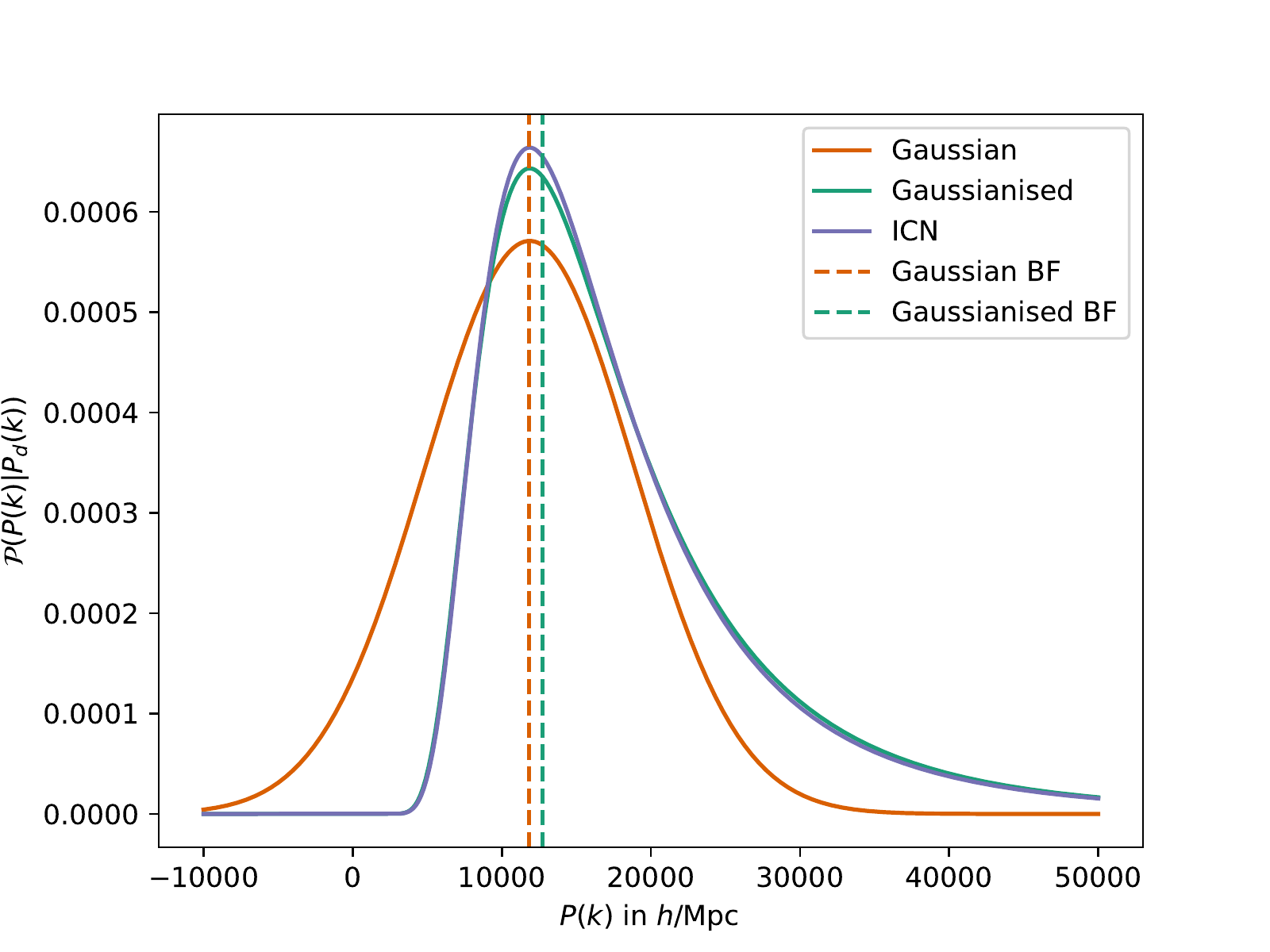}
    \caption{Comparison of a Gaussian posterior distribution for the power spectrum $P(k)$ (red) for the lowest $k$-bin centred around $k=0.0014\;h/\mathrm{Mpc}$ with a Gaussian posterior in the Gaussianised observable $\Psi(k)$ (green) and the inverse cubic normal distribution (ICN, purple, \cite{Kalus:2015lna}) as an estimate of the $P(k)$-distribution assuming Gaussianity in the density field $\delta(k)$. The vertical dashed-lines indicate the position of the model power spectra evaluated at the best-fitting cosmological parameters assuming a Gaussian $P(k)$ (orange) or a Gaussianised $\Psi(k)$ (green). The corresponding parameter shifts are listed in Table \ref{tab:FS_shifts_G_vs_nG}. The bulk-motion dipole has been set to the one measured by Planck.}
    \label{fig:Gauss_vs_nonGauss}
\end{figure}

Here we motivate our choice of gaussianized variables  of eq. \eqref{eq:GaussianisedPower} for the Fisher-based approach;   
in Table \ref{tab:FS_shifts_G_vs_nG}, we compare the results from using directly  $P(k)$ to those obtained using the Gaussianised power spectrum $\Psi(k)$. Considering the non-Gaussian distribution of the power spectrum at ultra-large scales increases the systematic shift due to the Kaiser rocket effect by about a factor of four. In the case that the NVSS dipole describes our motion, we would even disfavour $f_\mathrm{NL}=0$ at 1.7$\sigma$. To understand this dramatic increase in the bias, we have to keep in mind that the largest scales contribute the most to the shift in $f_\mathrm{NL}$. For that reason, we compare in Figure \ref{fig:Gauss_vs_nonGauss} a Gaussian distribution  the power spectrum value $P(k)$ at the lowest $k$-in at $k=0.0014\;h/\mathrm{Mpc}$ with the one using the Gaussianised observable $\Psi(k)$ and the inverse cubic normal distribution \cite[ICN]{Kalus:2015lna} that has been shown to  approximate very well the power spectrum distribution for a Gaussian density field $\delta(k)$. The effective number of modes $M$ (cf. eq. \eqref{eq:modes}) is 5.8 and, therefore, assuming a Gaussian distribution for the power spectrum gives unphysical results, such as a non-zero probability of the power spectrum being negative. Evaluating our power spectrum model for the best-fitting cosmological parameters obtained assuming a Gaussian $P(k)$ provides a value that is close to the peak in all distributions plotted in Figure \ref{fig:Gauss_vs_nonGauss}. However, as the true distribution is suppressed towards small power spectrum values and extends into a tail at higher values, the model power spectrum value established from shifted best-fitting parameter values from the Gaussianised observable $\Psi$ is closer to the mean of the true distribution. This justifies our choice in the main text.

\begin{table}[ht]
    \centering
    \begin{tabular}{|l|c|c|c|c|c|c|}
      \hline summary &  $\Delta_{\rm sys}$&$\Delta_{\rm sys}$&$\Delta_{\rm sys}$&$\Delta_{\rm sys}$&$\Delta_{\rm sys}$&$\Delta_{\rm sys}$\\
          statistic & $h$ & $\omega_\mathrm{b}$ & $\omega_\mathrm{cdm}$ & $n_s$ & $b$ & $f_\mathrm{NL}$ \\\hline\hline
         $P(k)$  &  -9.0$\times 10^{-7}$
 &  -1.5$\times 10^{-7}$
 &  5.0$\times 10^{-6}$
 &  7.0$\times 10^{-6}$
 &  -6.8$\times 10^{-5}$
 &  0.29\\
         &  -0.00037 $\sigma$
 &  -0.0010 $\sigma$
 &  0.0055 $\sigma$
 &  0.0023 $\sigma$
 &  -0.0046 $\sigma$
 &  0.027 $\sigma$\\
         $\Psi(k)$ &  1.2$\times 10^{-5}$
 &  -2.4$\times 10^{-7}$
 &  1.1$\times 10^{-5}$
 &  2.8$\times 10^{-5}$
 &  -0.00028
 &  1.1
\\
          & 0.0048 $\sigma$
 &  -0.0016 $\sigma$
 &  0.012 $\sigma$
 &  0.0092 $\sigma$
 &  -0.019 $\sigma$
 &  0.10 $\sigma$\\\hline
    \end{tabular}
    \caption{Expected shift of the best-fitting $\Lambda$CDM+$f_\mathrm{NL}$ parameters similar to Table \ref{tab:FS_shifts_main} but comparing results obtained directly using as observable directly the power spectrum $P(k)$, i.e. ignoring the non-Gaussian distribution of the power spectrum at the most relevant scales, and using a Gaussianised version $\Psi(k)$ of the galaxy power spectrum.}
    \label{tab:FS_shifts_G_vs_nG}
\end{table}

\acknowledgments
We thank the anonymous referee, as well as Enzo Branchini, David Parkinson, Dominik Schwarz and Roy Maartens for useful comments and feedback  on earlier versions of this manuscript. We express our gratitude to H\'ector Gil-Mar\'in for providing us with code to produce Gaussian random field realisations. We made use of Matplotlib \cite{Hunter:2007} for plotting. Other software packages that we used include  We use Martin Krzywinski's colourblind-friendly colour palette.\footnote{http://mkweb.bcgsc.ca/colorblind/palettes/8.color.blindness.palette.txt} Some of the results in this paper have been derived using the healpy and HEALPix packages \cite{Gorski:2004by,Zonca2019}.\footnote{http://healpix.sourceforge.net} 

BBK is supported by
the project 우주거대구조를 이용한 암흑우주 연구 ("Understanding
Dark Universe Using Large Scale Structure of
the Universe"), funded by the Korean Ministry of Science, and has been supported by the European Union’s
Horizon 2020 research and innovation programme ERC
(BePreSySe, grant agreement 725327), and Spanish MINECO
under project PGC2018-098866-B-I00  FEDER, UE.
DB acknowledges partial financial support by ASI Grant No. 2016-24-H.0 and funding from Italian Ministry of Education, University and Research (MIUR) through the "Dipartimenti di eccellenza project Science of the Universe.

\bibliographystyle{JHEP}
\bibliography{Rocket}{}
\end{document}